\documentclass[sigchi, table]{acmart}
\usepackage{booktabs}
\usepackage{array}
\usepackage{caption}
\usepackage{graphicx}
\usepackage{siunitx}
\usepackage{xcolor}
\usepackage{multirow}
\usepackage{hhline}
\usepackage{calc}
\usepackage{tabularx}

\usepackage{balance}
\usepackage[inline]{enumitem}
\usepackage{sparklines}

\copyrightyear{2019} 
\acmYear{2019} 
\setcopyright{acmlicensed}
\acmConference[CHI 2019]{CHI Conference on Human Factors in Computing Systems Proceedings}{May 4--9, 2019}{Glasgow, Scotland UK}
\acmBooktitle{CHI Conference on Human Factors in Computing Systems Proceedings (CHI 2019), May 4--9, 2019, Glasgow, Scotland UK}
\acmPrice{15.00}
\acmDOI{10.1145/3290605.3300297}
\acmISBN{978-1-4503-5970-2/19/05}

\begin{document}
\title{When Do People Trust Their Social Groups?}

\author{Xiao Ma$^1\dagger$, Justin Cheng$^2$, Shankar Iyer$^2$, Mor Naaman$^1$}
\email{{xiao,mor}@jacobs.cornell.edu, {jcheng,shankar94}@fb.com}
\renewcommand{\shortauthors}{Xiao Ma, Justin Cheng, Shankar Iyer, Mor Naaman}
\affiliation{%
  \institution{$^1$Jacobs Institute, Cornell Tech, $^2$Facebook}
  {\small {$^\dagger$ Work done while at Facebook.}}
}

\begin{abstract}

Trust facilitates cooperation and supports positive outcomes in social groups, including member satisfaction, information sharing, and task performance. 
Extensive prior research has examined individuals' general propensity to trust, as well as the factors that contribute to their trust in specific groups.
Here, we build on past work to present a comprehensive framework for predicting trust in groups.
By surveying 6,383 Facebook Groups users about their trust attitudes and examining aggregated behavioral and demographic data for these individuals, we show that (1) an individual's propensity to trust is associated with how they trust their groups, (2) smaller, closed, older, more exclusive, or more homogeneous groups are trusted more, and (3) a group's overall friendship-network structure and an individual's position within that structure can also predict trust.
Last, we demonstrate how group trust predicts outcomes at both individual and group level such as the formation of new friendship ties.

\end{abstract}
 



\maketitle

\section{Introduction}

Trust contributes to the success of social groups by encouraging people to interpret others' actions and intentions favorably, thereby facilitating cooperation and a sense of community \cite{gambetta1988trust,misztal2013trust,uzzi1996sources,bachmann2001trust,dirks1999effects,preece2005online}.
In groups, trust increases member satisfaction and task performance \cite{walther2005rules}, reduces conflict~\cite{gambetta1988trust,walther2005rules}, and promotes effective response to crisis~\cite{meyerson1996swift}.

Previous research has examined how different factors such as  size~\cite{brewer1991social,denters2002size,zelmer2003linear}, group cohesiveness~\cite{hogg1993group}, and activity~\cite{walther2005rules} may impact people's trust in their social groups, both online~\cite{holtz2017social} and offline~\cite{rotter1971generalized}.
However, previous studies tend to be small in scale, limited to specific contexts (e.g., online marketplaces),
or only consider a specific type of group (e.g., organizations~\cite{mayer1995integrative,colquitt2007trust}).
Studies that address these three limitations may enrich our understanding of how trust is formed in social groups more generally.

In this work, we build on this rich prior literature on trust to present a framework for predicting an individual's trust in a social group, and examine how differences at the individual and group levels predict that trust.
We focus our analysis on Facebook Groups, a Facebook feature that ``allows people to come together to communicate about shared interests''~\cite{facebook2018help}.
As of May 2018, 1.4 billion people used Facebook Groups every month \cite{facebook2018launch}.
By combining a survey ($N$=$6,383$ valid responses) of individuals using Facebook Groups with aggregated behavioral logs, we are able to investigate, across a diverse sample, how an individual's trust in a group relates to characteristics of the individual, the group, and the individual's membership in that group.

The survey asked individuals about their general attitudes towards others and trust towards a Facebook group that they were a member of.
While prior work has shown that an individual's general propensity to trust others influences their trust in a particular group \cite{boss1978trust, butler1999trust, ridings2002some, ferguson2015sinking}, we additionally examine the role of other individual-level differences (e.g, general attitudes towards risk-taking). 

We combine these survey results with aggregated behavioral and descriptive data on Facebook Groups. This allows us to study the role of five categories of features that characterize either the group or the respondent's relationship with the group, based on prior literature: (1) basic properties of the group (e.g., size, membership privacy policy)~\cite{kraut2014role}, (2) group category~\cite{denson2006roles}, (3) group activity~\cite{kraut2014role}, (4) group homogeneity~\cite{moser2017community}, and (5) the friendship-network structure of the group ~\cite{holtz2017social}.

We find that these variables robustly predict participants' trust in a particular group, with both individual and group characteristics contributing predictive value (adjusted $R^2$=$0.26$).
In particular, an individual's trust in a group was most strongly predicted by their general perceived social support, the group's average clustering coefficient, and their degree centrality in the group.
We also show that trust in a group can be estimated using only observational data.

While these results support previous findings showing that intragroup trust decreases with increasing group size and increases with membership restriction~\cite{zelmer2003linear,denters2002size,brewer1991social,la2016small,moser2017community}, we find that these trends only hold up to a certain point.
When the size of a group exceeds 150 members (roughly Dunbar's number, or the expected cognitive limit beyond which social relationships are difficult to maintain \cite{dunbar1992neocortex}), the membership policy of the group (public v.s. private) ceases to play a predictive role.
Moreover, in deciding how much to trust a group, we show that group size matters less to individuals with a higher general propensity to trust.

Further, previous work suggests that people trust groups in which they are more active \cite{cartwright1953group}, but we find that only certain types of activities are associated with trust:
people ``like'' and ``comment'' more in groups they trust but do not necessarily post more, suggesting that trust is associated more with directed communication than with information sharing.

Finally, we show that trust in groups is associated with both individual- and group-level outcomes.
Increased trust leads to individuals being more likely to form friendships with other members of the group, but is also associated with the group being less likely to grow larger in size.

In summary, we
\begin{enumerate*}
    \item present results of a large survey of individuals' trust attitudes towards their social groups,
    \item examine how characteristics of both the individual and group contribute to trust in a group, and
    \item show how this trust affects future individual- and group-level outcomes.
\end{enumerate*}
A deeper understanding of how these factors collectively contribute to trust in groups can better equip communities to foster trust among their members.

\section{Background}
To begin, we describe previous work on trust and the factors that impact it. This prior work also motivates the multilevel approach that we take for studying trust in social groups.

\subsection{What is Trust?}
Trust has been defined as a belief in the reliability of others~\cite{gambetta1988trust}. 
Previous work on trust can be roughly organized into examining trust
\begin{enumerate*}
    \item as a personal trait (i.e., a propensity to trust others),
    \item with respect to a specific other individual (i.e., dyadic trust), or
    \item with respect to multiple others (e.g., in groups or organizations).
\end{enumerate*}
These differences roughly correspond to how trust is conceptualized across disciplines---as arising from individual traits in psychology \cite{rotter1971generalized}, as calculable using game theory in economics \cite{williamson1993calculativeness}, or from relationships among people in sociology \cite{granovetter1985economic}.
In this broad literature, trust has typically been either measured using surveys \cite{wvs, rotter1971generalized}, qualitative interviews \cite{six2008creating}, or through economic games that measure how much money people entrust others with \cite{berg1995trust}.
In the present work, we measure trust using a survey asking about trust and its various correlates, and combine this with observed behavioral data.

\subsubsection{Trust and the individual}
One line of work has studied trust as an individual characteristic, similar to a personality trait, and suggests that trust is %
rooted in life experiences and societal norms~\cite{rotter1971generalized,bowlby1969attachment,ainsworth2015patterns}.
In this context, trust is referred to as ``generalized trust'' \cite{nannestad2008have}, a ``propensity to trust''~\cite{ferguson2015sinking}, or a ``disposition to trust''~\cite{wu2010effects}.
A propensity to trust others has been associated with being older, married, having higher family income and college education and living in a rural area, but not with gender~\cite{taylor2007americans, paxton2007association}.
Work has also studied cross-country differences in an individual's propensity to trust~\cite{bjornskov2007determinants}. %
In this work, we refer to this ``individual trait'' definition of trust as a ``disposition to trust'', which we measure in order to explain trust in groups.
To minimize cross-cultural effects, we focus on U.S.-based individuals.
A disposition to trust is also related to other personal traits, such as risk-taking ~\cite{cook2005trust}, feelings of social support, and group loyalty~\cite{barrera1983structure, van2004social}.
We include these relevant concepts in our work. %

\subsubsection{Trust between individuals}
At a dyadic level, trust can be modeled using social exchange theory \cite{berg1995trust,homans1958social,blau1964power,emerson1976social}, where
people are assumed to be rational actors who maximize their own benefits in interactions with others.
Trust is then defined as the decision to undertake risks in these exchanges.
Another significant line of work has examined dyadic trust in online settings.
This work suggests that reputation ~\cite{yamagishi2004solving,resnick2002trust,kuwabara2015reputation,qiu2018more}, homophily \cite{abrahao2017reputation}, and the language used in online profiles \cite{ma2017self} mediate an individual's trust in someone else.

\subsubsection{Trust in groups and organizations}
While it is useful to study social interactions at dyadic level, many interactions take place in the context of groups, both offline and online. 
Significant work has studied how trust influences an organization's structure, productivity, and cohesiveness \cite{mayer1995integrative,nyhan2000changing,mcevily2003trust,fine1996secrecy}.
Trust in organizations was positively associated with improved job performance, citizenship behavior (e.g., altruism and courtesy), and greater cooperation. It is negatively associated with counterproductive activities such as disciplinary action and tardiness \cite{dirks1999effects,colquitt2007trust}.
Trust in online groups impacts various outcomes including member satisfaction, information sharing, and task performance in virtual teams \cite{walther2005rules}, but is also shown to be fragile and temporal when the team is formed around a common task with a finite life span~\cite{jarvenpaa1999communication,meyerson1996swift}. 

More recently, several studies have looked at trust in Facebook groups, mostly in the context of buy-and-sell groups~\cite{holtz2017social,moser2017community}. 
Qualitative research on Facebook buy-and-sell groups showed that trust can be fostered through mechanisms such as closed membership and sanctioning ~\cite{moser2017community}, and our quantitative analysis here confirms some of these findings.

\subsubsection{Multilevel perspectives on trust}
Given the diversity of approaches used to study trust, some work has called for ``multilevel perspectives'' on trust \cite{rousseau1998not} that account for trust at the individual, group, and organizational levels.
Research has proposed models of how trust in others depends on a disposition to trust \cite{mayer1995integrative}.
And as previously noted, studies have also examined how trust at the individual level may mediate trust at other levels (e.g., in a specific community \cite{ridings2002some} or in groups \cite{boss1978trust, ferguson2015sinking}).
Nonetheless, such work remains largely theoretical.
Empirical studies have tended to be small-scale and only examine a subset of the many characteristics and behaviors of individuals and groups that may mediate trust.
In this work, we show how trust may be better modeled by considering individual and group-level features together in a generalizable framework. %
\subsection{Determinants of Trust in Groups}
What contributes to trust in groups?
Here we review relevant literature that guide the selection of our feature sets in predicting trust in groups.

\subsubsection{Individual differences}
Trust in groups can be mediated by one's disposition to trust others, as it correlates with one's initial intentions to trust a group, especially in ambiguous situations~\cite{gill2005antecedents}. 
A disposition to trust can positively impact trust in different settings, including trust between individuals~\cite{yakovleva2010we}, in communities~\cite{ridings2002some}, in organizations~\cite{kantsperger2010consumer}, or in online services~\cite{wu2010effects}. 
Similarly, a disposition to trust increases trustworthiness evaluations given to Airbnb hosts~\cite{ma2017self}, though in other settings, a disposition to trust was not associated with trust in peer sellers~\cite{jones2008trust}. 

Past work also suggests an inverse relationship between risk aversion and trust~\cite{abrahao2017reputation}---the more comfortable an individual is with taking risks, the higher the trust they have in groups.

Further, prior literature treats membership of voluntary associations as an indicator of trust~\cite{putnam2000bowling,putnam1993prosperous}. %
Thus, greater
in-group loyalty, as well as perceived social support from others, should both be linked with higher trust in groups due to increased group participation and perceived social capital.

\subsubsection{Group characteristics}
Trust in groups may also stem from basic properties of the group such as its size~\cite{brewer1991social,denters2002size,zelmer2003linear}.
For instance, experiments have shown that people identify more strongly with smaller groups \cite{simon1987perceived}.
Older groups should also be trusted more, as they have more time to mature in norm and management.
Past research has also described how secrecy can build community \cite{fine1996secrecy} and shown that group cohesiveness promotes trust~\cite{stokes1983components}.
Recent qualitative work on Facebook Groups also suggests that by making membership exclusive and screening new members, trust can be enhanced~\cite{moser2017community}. 

Homogeneity, which relates to cohesiveness, may also contribute to trust.
People tend to be closer to and trust others who are similar to them~\cite{mcpherson2001birds}. 
Research has also found a relationship between gender and age homophily and increased trust \cite{ahmad2011trust,abrahao2017reputation}.

Higher levels of group activity are also linked with greater trust~\cite{cartwright1953group,walther2005rules}.
Increased social interaction provides ``opportunities for people to get acquainted, to become familiar with one another, and to build trust''~\cite{ren2007applying}, thus leading to higher familiarity, and in turn, greater trust \cite{gulati1995does}.

\subsubsection{Network characteristics}
Beyond group characteristics mentioned above, the overall structure of relationships between individuals in the group, as well as the individual's embeddedness the group's social network may mediate trust. 
A person's number of friends and the connections among these friends can both increase the likelihood of them joining a community~\cite{backstrom2006group, ugander2012structural}. 
As dense networks promote cooperation and social norms, they are also likely to be associated with increased trust~\cite{coleman1988social}.  
In buy-and-sell groups on Facebook, network density and the degree centrality of the seller are positively correlated with an intention to transact, which may signal higher trust in the group~\cite{holtz2017social}.
Our work uses similar features but directly measures trust via a survey, and considers the role of network features within a much large set of variables. %

This rich prior literature motivates our analysis in this work, in which we conduct a large-scale survey and analyze behavioral data to show how individual- and group-level differences help predict trust in groups.
Our research questions are as follows: (a) Can a baseline model that accounts only for individual attitudes predict trust in groups? (b) What is the relative contribution of the different sets of group-level features (basic group properties, group category, activity, homogeneity, and structural properties) on trust in groups beyond the baseline model?
\section{Methods}

In this work, we conducted a survey of 10,000 respondents to a random sample of active Facebook Groups users in the U.S.
People were invited to participate in the survey via an ad on Facebook.
The survey was designed to measure both individual attitudes as well as trust in one of the randomly selected Facebook groups of which they were members.
We augmented this survey data with self-reported demographic data such as age and gender and server logs of these individuals' activity and friendships on Facebook.
All log data was de-identified and analyzed in aggregate on Facebook's servers; researchers did not view any identifiable data nor access any specific posts in any groups.
The study was approved by an internal Facebook board as well as Cornell's Institutional Review Board under protocol \#1805008006. 

\subsection{Sampling}
The survey was issued to unique individual-group pairs.
We used the following sampling strategy to identify eligible survey candidates.
First, we identified Facebook groups that had at least five members and that had a majority of their members located in the U.S.
We then identified people in the U.S. who belong to at least one of these groups, and that had at least one interaction (e.g., creating, liking, or commenting on a group post) in the past 28 days.
We then sampled eligible individual-group pairs, de-duplicating by both individual and group at random.
The sampling was also stratified by group size (the number of members in the group) to better capture behavior across both smaller and larger groups.
We note the following bias introduced by our sampling method: compared to all individuals who actively used groups in the past 28 days, our participants tended to be 8.7\% older and were 17.5\% more likely to be women. 

\subsection{Survey Design}
The survey consisted of two sections: a section on individual differences regarding the participant's general attitudes towards others, including disposition to trust and related concepts; and a section on trust in a specific Facebook group.
Each section had four items, shown in~\autoref{tab:trust_survey}.
The order of questions was randomized within a section.
Participants were asked to report the extent to which they agreed or disagreed with each statement on a five-point Likert scale.

For the section on general attitudes towards others, we measured disposition to trust through an adaptation of the generalized trust question in the World Values Survey~\cite{wvs}.
The original question elicits a dichotomous response, worded as: ``Generally speaking, would you say that most people can be trusted or that you need to be very careful in dealing with people?''
We instead used a more granular five-point Likert scale, which has been shown to be more reliable~\cite{miller2003surveys}.
We also included measures of concepts related to disposition to trust reported in previous literature, including general social support~\cite{barrera1983structure,vigoda2010organizational,hether2014s}, in-group loyalty~\cite{van2004social}, and risk aversion~\cite{miller2003surveys}.

To measure an individual's level of trust in a Facebook group of which they were a member, we created a four-item scale to measure trust in groups (section two in \autoref{tab:trust_survey}), based on previous literature.
This scale is based on the framework of ability, integrity, and benevolence by Mayer et al.~\cite{mayer1995integrative} and Schorman et al.~\cite{schoorman2007integrative}, and also adapts measures from several interpersonal trust scales including Rotter Interpersonal Trust Scale~\cite{rotter1971generalized}, the Specific Interpersonal Trust Scale~\cite{johnson1982measurement}, and a newer ``predisposition to trust'' scale~\cite{ashleigh2012new}.

In addition, to better understand what people use the group for, we also asked participants to use the taxonomy below to describe the group category:

\begin{itemize}
    \item Friends \& Family: e.g., close friends, extended family
    \item Education \& Work: e.g., college, job, professional
    \item Interest-Based: e.g., hobby, book club, sports
    \item Identity-Based: e.g., lifestyle, health, faith, parenting
    \item Location-Based: e.g., neighborhood or local organization
    \item Other
\end{itemize}

These categories were identified in previous
qualitative research, where we surveyed people who used Facebook Groups and asked them to describe a group they were part of (e.g., ``my family'').
In our work, participants were requested to select all categories that applied to the group they were surveyed on, and we treated each group category as a binary variable.
In our sample, around 34\% of the groups were tagged as interest-based groups (most common), followed by 20\% friends \& family groups.
The first five categories capture most of the groups (covering 89\%).

\begin{table}
\footnotesize
\begin{tabular}[0.95\textwidth]{p{.17\linewidth}p{.1\linewidth}p{.63\linewidth}}
\toprule
\multicolumn{3}{l}{\textbf{General Attitudes Towards Others}} \\
\midrule
\multicolumn{2}{l}{Disposition to trust} & Most people can be trusted. \\
\multicolumn{2}{l}{General social support} & There are people in my life who give me support and encouragement. \\
\multicolumn{2}{l}{General risk attitude} & I'm willing to take risks. \\
\multicolumn{2}{l}{General in-group loyalty} & I would describe myself as a ``team player''. \\
\midrule
\multicolumn{3}{l}{\textbf{Trust in a Group}} \\
\midrule
Care & \multicolumn{2}{p{.73\linewidth}}{Other members of the group care about my well-being.}  \\
Reliability & \multicolumn{2}{p{.73\linewidth}} {Other members of this group can be relied upon to do what they say they will do.} \\
Integrity & \multicolumn{2}{p{.73\linewidth}}{Other members of this group are honest.} \\
Risk-taking  & \multicolumn{2}{p{.73\linewidth}}{I feel comfortable sharing my thoughts in this group.} \\
\bottomrule
\end{tabular}
\caption{Trust in groups survey items. Participants reported the degree to which they agreed or disagreed to each of the survey items on a five-point Likert scale.}
\label{tab:trust_survey}
\end{table} 
\subsection{Data and Statistical Approaches}
In addition to data from the survey, we examined properties of groups including their sizes and membership privacy policies.
For each group, we also looked at an individual's activity in the group (e.g., time spent, likes, comments, and posts made), the group's overall activity, as well as group members' friendships with each other.

Out of the 10,000 survey responses we received, we filtered responses based on the completeness of the survey, as well as availability of self-reported and log data.
In the end, we retained 6,383 responses for our main analysis.

The main statistical techniques we used were multiple linear regression, random forests, and logistic regression.
We first built a baseline model predicting trust in groups using variables capturing individual-level differences.
Then, we identified five different sets of group-level features, conducted analysis on how much each set of feature improved the baseline model, and interpreted the relationship between each feature and trust separately.
We next combined all features in a random forest model and compared the importance of each set of features in the combined model.
Finally, we used logistic regression to predict group outcomes such as the densification of the friendship network within the group.
When appropriate, we log-transformed the data (e.g., group size) and note the transformation when reporting coefficients.

\section{Results}

\definecolor{16gray}{gray}{0.98}
\definecolor{19gray}{gray}{0.96}
\definecolor{20gray}{gray}{0.95}
\definecolor{24gray}{gray}{0.91}
\definecolor{25gray}{gray}{0.90}
\definecolor{38gray}{gray}{0.77}

\definecolor{54gray}{gray}{0.71}
\definecolor{56gray}{gray}{0.69}
\definecolor{59gray}{gray}{0.66}
\definecolor{60gray}{gray}{0.65}
\definecolor{62gray}{gray}{0.63}
\definecolor{67gray}{gray}{0.58}

\begin{table}
\footnotesize
\begin{tabular}[.95\linewidth]{p{.46\linewidth}p{.05\linewidth}p{.04\linewidth}p{.07\linewidth}p{.07\linewidth}p{.07\linewidth}}
\toprule
\textbf{Variable}           & \textbf{Mean} & \textbf{SD}   &  \multicolumn{3}{l}{\textbf{Correlations}} \\ 
                            &      &      & 1         & 2         & 3        \\
\midrule
\multicolumn{6}{l}{\textbf{General Attitudes Towards Others}} \\
\midrule
1. Disposition to trust   \hspace{.2em}    
\begin{sparkline}{5}
    \sparkspike .0 .13
    \sparkspike .15 .34
    \sparkspike .3 .51
    \sparkspike .45 .82
    \sparkspike .6 .20
\end{sparkline}
& 3.33 & 1.07 &           &           &          \\
2. General risk attitude    \hspace{.2em} 
\begin{sparkline}{5}
    \sparkspike .0 .08
    \sparkspike .15 .21
    \sparkspike .3 .42
    \sparkspike .45 .94
    \sparkspike .6 .36
\end{sparkline} 
& 3.68 & 0.99 & 0.18*** \cellcolor{20gray}    &           &          \\
3. General in-group loyalty  \hspace{.2em} 
\begin{sparkline}{5}
    \sparkspike .0 .03
    \sparkspike .15 .05
    \sparkspike .3 .22
    \sparkspike .45 .68
    \sparkspike .6 1.02
\end{sparkline}
& 4.54 & 0.80 & 0.23*** \cellcolor{25gray}   & 0.16*** \cellcolor{16gray}   &          \\
4. General social support  \hspace{.2em} 
\begin{sparkline}{5}
    \sparkspike .0 .03
    \sparkspike .15 .03
    \sparkspike .3 .13
    \sparkspike .45 .48
    \sparkspike .6 1.32
\end{sparkline}  
& 4.34 & 0.84 & 0.24***  \cellcolor{24gray}  & 0.18***  \cellcolor{20gray}  & 0.36***  \cellcolor{38gray} \\ 
\midrule
\multicolumn{6}{l}{\textbf{Trust in Groups}} \\
\midrule
1. Care \hspace{.2em}    
\begin{sparkline}{5}
    \sparkspike .0 .08
    \sparkspike .15 .07
    \sparkspike .3 .59
    \sparkspike .45 .50
    \sparkspike .6 0.76
\end{sparkline}
& 3.90 & 1.08 &           &           &          \\
2. Reliability  \hspace{.2em}    
\begin{sparkline}{5}
    \sparkspike .0 .04
    \sparkspike .15 .05
    \sparkspike .3 .56
    \sparkspike .45 .50
    \sparkspike .6 0.86
\end{sparkline}
& 4.05 & 0.98 & 0.62*** \cellcolor{62gray}  &           &          \\
3. Integrity   \hspace{.2em}    
\begin{sparkline}{5}
    \sparkspike .0 .03
    \sparkspike .15 .03
    \sparkspike .3 .46
    \sparkspike .45 .47
    \sparkspike .6 1.02
\end{sparkline}
& 4.20 & 0.95 & 0.60*** \cellcolor{60gray}  & 0.67*** \cellcolor{67gray}   &          \\
4. Risk-taking \hspace{.2em}    
\begin{sparkline}{5}
    \sparkspike .0 .07
    \sparkspike .15 .08
    \sparkspike .3 .37
    \sparkspike .45 .55
    \sparkspike .6 0.92
\end{sparkline}
& 4.09 & 1.06 & 0.59*** \cellcolor{59gray}   & 0.54***  \cellcolor{54gray}  & 0.56***  \cellcolor{56gray} \\
\bottomrule
\multicolumn{6}{l}{\textit{Note:} $^{*}$p$<$0.05; $^{**}$p$<$0.01; $^{***}$p$<$0.001} \\ 
\end{tabular}
\caption{Descriptive summary of survey measures, including general attitudes and trust in groups. Sparklines represent the histogram of each measure. (N=6,383)}
\label{tab:survey_measure_desc}
\end{table} 
Trust in groups was measured in our survey across four dimensions: care, reliability, integrity, and risk taking.
As shown in \autoref{tab:survey_measure_desc}, these dimensions of trust in groups are highly correlated ($\rho \ge$ 0.54; Cronbach's $\alpha$ = 0.86).
Thus, we defined a composite ``trust in groups'' score as the mean of all four dimensions and report findings with respect to this composite score.

\subsection{Individual Differences and Trust}
We start by predicting trust in groups using individual attitudes as well as demographic information (see in~\autoref{tab:baseline}), which prior literature has associated with differences in one's disposition to trust~\cite{taylor2007americans}. %

\begin{table} 
\footnotesize
\centering 
\begin{tabular}[0.95\linewidth]{p{0.28\linewidth}p{0.08\linewidth}p{0.08\linewidth}p{0.08\linewidth}p{0.08\linewidth}p{0.08\linewidth}}
\toprule
 & \multicolumn{5}{c}{\textit{Dependent variable:}} \\ 
\cline{2-6} 
\\[-1.8ex] & \multicolumn{5}{c}{Trust in groups composite score} \\ 
\\[-1.8ex] & (1) & (2) & (3) & (4) & (5)\\ 
\midrule
 (Intercept) & 4.07$^{***}$ & 3.51$^{***}$ & 3.18$^{***}$ & 2.46$^{***}$ & 1.98$^{***}$ \\ 
  & (0.04) & (0.05) & (0.06) & (0.07) & (0.08) \\ 
  Age & $-$0.001$^{*}$ & $-$0.003$^{***}$ & $-$0.002$^{**}$ & $-$0.001$^{*}$ & $-$0.001$^{*}$ \\ 
  & (0.001) & (0.001) & (0.001) & (0.001) & (0.001) \\ 
 Female & 0.09$^{***}$ & 0.06$^{**}$ & 0.08$^{***}$ & 0.05$^{*}$ & 0.02 \\ 
  & (0.02) & (0.02) & (0.02) & (0.02) & (0.02) \\ 
 Disposition to trust &  & 0.19$^{***}$ & 0.17$^{***}$ & 0.13$^{***}$ & 0.11$^{***}$ \\ 
  &  & (0.01) & (0.01) & (0.01) & (0.01) \\ 
 Risk attitude &  &  & 0.09$^{***}$ & 0.07$^{***}$ & 0.05$^{***}$ \\ 
  &  &  & (0.01) & (0.01) & (0.01) \\ 
 In-group loyalty &  &  &  & 0.21$^{***}$ & 0.16$^{***}$ \\ 
  &  &  &  & (0.01) & (0.01) \\ 
 Social support &  &  &  &  & 0.19$^{***}$ \\ 
  &  &  &  &  & (0.01) \\ 
\midrule
Adjusted R$^{2}$ & 0.003 & 0.06 & 0.07 & 0.11 & 0.14 \\ 
\bottomrule
\multicolumn{5}{l}{\textit{Note:} $^{*}$p$<$.05; $^{**}$p$<$.01; $^{***}$p$<$.001} \\ 
\end{tabular} 
  \caption{Baseline model predicting trust in groups using demographics, disposition to trust, risk attitude, in-group loyalty, and social support. (N=6,323 after removing missing age and gender observations)} 
  \label{tab:baseline} 
\end{table}

\subsubsection{Demographics}
We found that demographic factors such as the age and gender of participants capture almost no variance of trust in groups (see Model~1 in~\autoref{tab:baseline}).
This result partially contrasts with the prior work that found a relationship between these demographic factors and one's disposition to trust~\cite{taylor2007americans}.
To better understand this result, we tested a model that used demographic variables to predict participants' disposition to trust rather than trust in groups.
While we found that older people were more trusting than young people ($\beta$=$0.006$, $p$<$.001$) and women were more trusting than men in general ($\beta$=$0.12$, $p$<$.001$), very little variance in disposition to trust is explained by these demographic factors [$R^2$=$0.01$, $F(2, 7174)$=$36.1$, $p$<$.001$].
In other words, demographic characteristics explain neither an individual's disposition to trust nor their trust in groups.

\subsubsection{General attitudes towards others}
How does an individual's general attitudes towards others predict their trust in groups?
Corroborating prior work, one's general disposition to trust significantly predicts one's trust in groups (see Model 2 in~\autoref{tab:baseline}).
However, other factors also play significant roles (Models 3--5 in~\autoref{tab:baseline}).
Notably, the individual's perceived social support ($\beta$=$0.19$, $p$<$.001$) and their general stated in-group loyalty ($\beta$=$0.16$, $p$<$.001$) contributed more to the prediction of trust in group than one's disposition to trust ($\beta$=$0.11$, $p$<$.001$).
A willingness to take risks ($\beta$=$0.05$, $p$<$.001$) was least predictive.
Altogether, these factors capture a significant amount of the variance in trust in groups (adjusted $R^2$=$0.14$).

\subsection{Group Differences and Trust}

To understand the relationship between group characteristics and trust in groups, we identified five distinct sets of group-level features (see \autoref{tab:group_features}).
In this section, we measure the incremental predictive value of each of these sets of group-level features, after controlling for the individual differences discussed above. Here, we use ``baseline model'' to refer to a model that only includes the individual differences (Model~5 in~\autoref{tab:baseline}). For each feature set, we add the features as independent variables in the multiple linear regression model to the baseline model. In each subsection, we report how much the model gains from the additional features. We validated that the coefficients of the individual differences features in the baseline do not change significantly when we include each new feature set.

\begin{table}
\footnotesize
\begin{tabular}[width=0.95\linewidth]{p{0.26\linewidth}p{0.69\linewidth}}
\toprule
\textbf{Feature Set} & \textbf{Features} \\ \midrule
Basic Properties (5)      & Group size, privacy type, group tenure, number of admins/moderators \\
Category (5)    & Self-reported group category \\
Activity (6)    & Group-level and participant-group-pair level time spent, number of posts, number of likes or comments \\
Homogeneity (3)  & Diversity of group age, gender, and similarity between participant and group average\\
Structural (5)  & Network density, average clustering coefficient, participant degree centrality, cliquishness of participant's friends in the group, average number of mutual friends with group members \\ \bottomrule
\end{tabular}
\caption{Five sets of group-level features used for predicting trust in groups.}
\label{tab:group_features}
\vspace{-.2in}
\end{table}

\subsubsection{Basic Group Properties}
The first set of group-level features consisted of group size, privacy type, group tenure~(how long a group has existed), the number of group admins, and its number of moderators. Adding these features to the baseline model increased the model's adjusted $R^2$ by $0.08$ ($p$<$.001$).
The most significant predictor of trust was group size.
Consistent with previous work on trust and group sizes~\cite{brewer1991social,denters2002size,zelmer2003linear}, people had lower trust in bigger groups ($\beta$=$-0.15$ on log scale, $p$<$.001$).

Apart from a group's size, 
a group's privacy type can also affect perceptions of trust.
On Facebook, group admins can set the group to be ``public'', ``closed'', or ``secret''.
Public groups are accessible to non-members, while closed and secret groups are only accessible to current members; closed groups differ from secret groups in whether their existence is known to non-members.
We found no significant differences between closed and secret groups, so we analyzed them together as ``private'' groups.

Controlling for group size (public groups are 68\% larger than private groups), we found that people trusted public groups \textit{less} than private groups  ($\beta$=$-0.07$, $p$<$.01$), as suggested in prior work \cite{moser2017community}.

Notably, we found an interaction effect between group size and privacy type in predicting trust ($\beta$=$0.04$, $p$<$.01$): the larger the group, the smaller the difference there is between trust in private and public groups.
To see how quickly this difference between group types dissipates, we conducted a series of t-tests in which we compared the mean difference in the trust composite score between public and private groups above a certain size threshold, starting from 10 in increments of 10.
These tests show significant differences between groups larger than the threshold until the threshold exceeds 150, where we no longer observe a significant difference between public and private groups ($p$>$.05$).

Incidentally, this size threshold roughly corresponds to Dunbar's number---the maximum number of stable social relationships a person can maintain due to limitations in cognitive resources~\cite{dunbar1992neocortex}. 
Smaller private groups provide control and exclusivity over membership, thus allowing members to foster a shared sense of identity~\cite{moser2017community}.
Once the group becomes too big, that shared identity might be lost, resulting in no difference between large groups that are public or private.

\begin{figure}
    \includegraphics[width=0.99\linewidth]{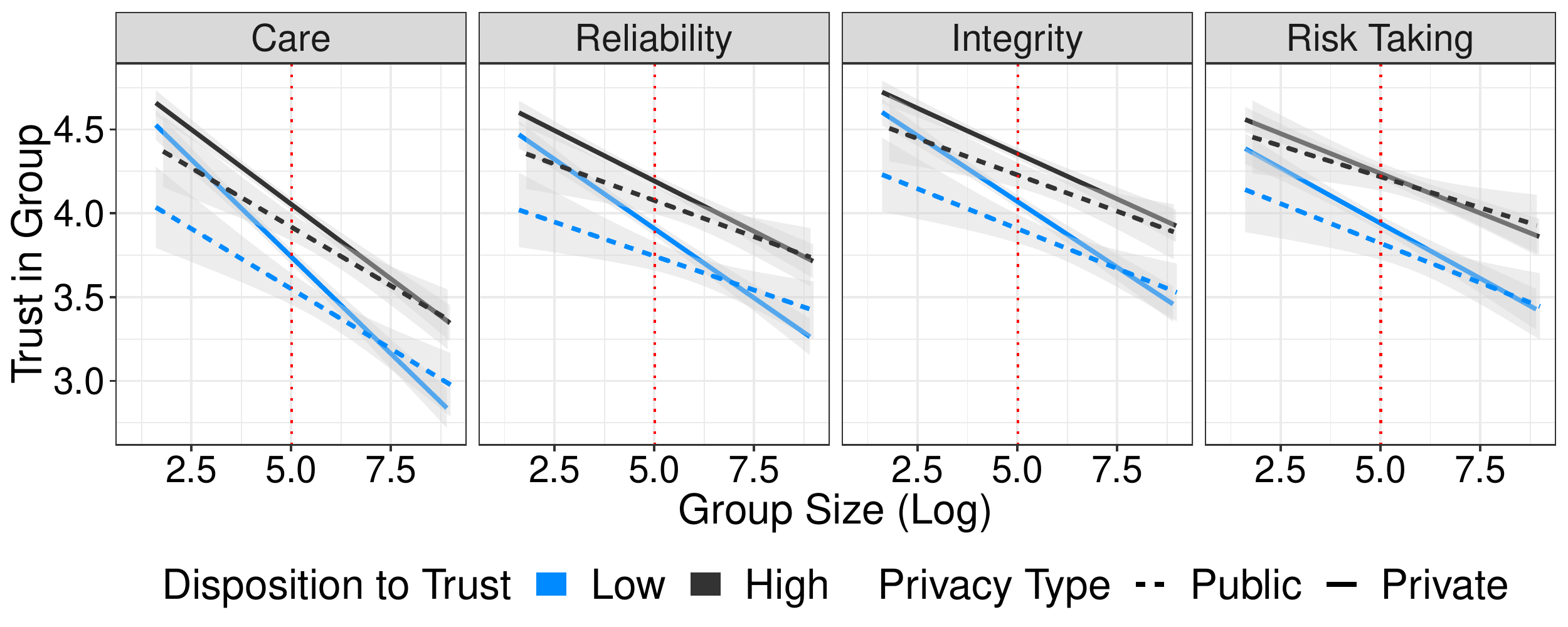}
    \caption{The relationship between trust in groups and group size, for each dimension (panels), across groups of different privacy types (line style) and individuals with different propensity to trust (line color). Dunbar's number (150) is marked by a vertical red dotted line.)
    }
    \label{fig:trust_in_groups_interaction}
\end{figure}

~\autoref{fig:trust_in_groups_interaction} summarizes the impact of group size on trust in public and private groups, as well as the effect of an individual's disposition to trust (we consider composite scores >3 to be high and <=3 to be low). The figure shows that having a high disposition to trust (black lines) and a group being private (solid lines) are both factors that contribute to trust in groups.
But while the effect of privacy decreases with size (dashed and solid lines cross), the reverse is true for an individual's disposition to trust.
An interaction effect between group size and individual's disposition to trust ($\beta$=$0.01$, $p$<$.01$) 
shows that people with a greater disposition to trust others were less sensitive to changes in group size (visually represented by gentler slope of black lines compared to blue ones in~\autoref{fig:trust_in_groups_interaction}).

Other basic group properties also relate to trust.
Longer group tenure predicts higher trust ($\beta$=$0.04$ on log scale, $p$<$.001$), potentially because older members have more stable group relationships and are more familiar with other group members~\cite{walther2005rules}.
The number of admins also predicts higher trust ($\beta$=$0.10$ on log scale, $p$<$.001$).
This finding is consistent with previous work that found that groups with more admins tended to survive longer than groups with fewer admins \cite{kraut2014role}.
The number of moderators is a much weaker predictor of group trust.

\subsubsection{Group category}
\begin{figure}
    \includegraphics[width=0.99\linewidth]{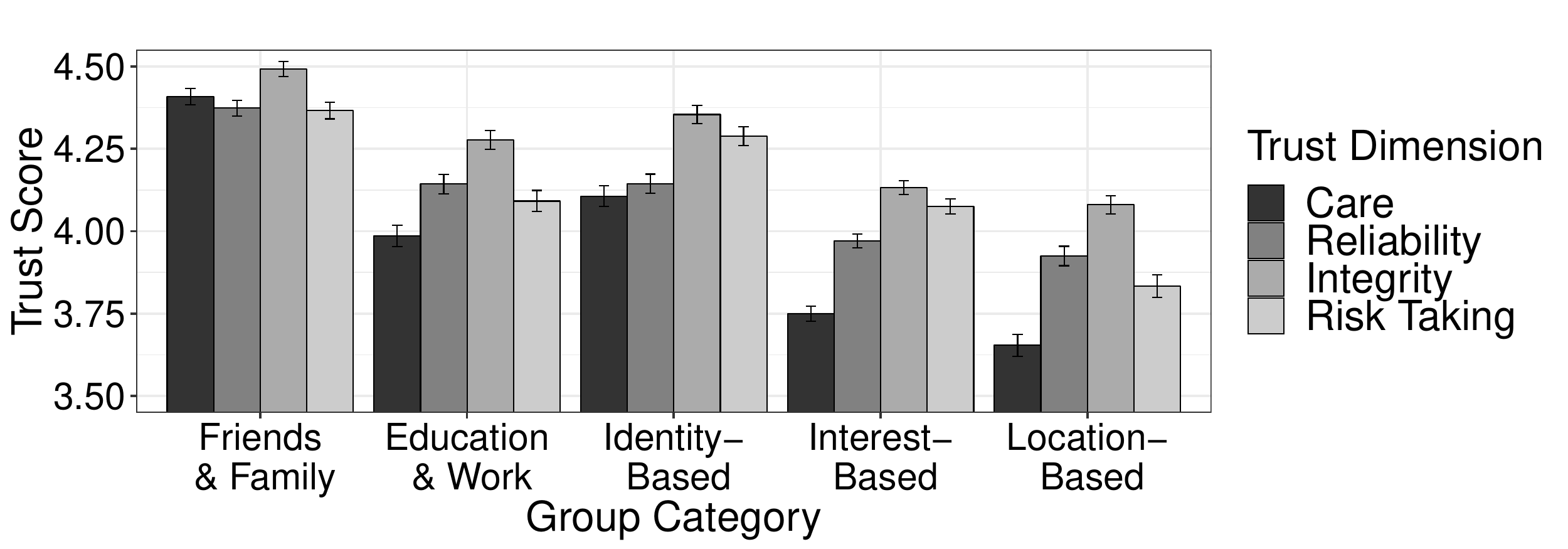}
    \caption{People have the highest trust in friends and family groups, and lowest in interest- and location-based groups.}
    \label{fig:trust_by_category}
\end{figure}

As previously described, participants in our survey labeled groups as belonging to one or more of six categories.
Including group category as multiple binary variables to the baseline model significantly improved trust predictions ($p$<$.001$), increasing the model's adjusted $R^2$ by $0.05$.
To illustrate differences in trust across these categories, we also conducted an ANOVA and plotted the average trust in groups by category in~\autoref{fig:trust_by_category}.
Groups marked as ``other'' were excluded from this analysis.
Post-hoc Tukey tests showed that people trust friends \& family groups the most, followed by identity-based and education \& work groups ($p$<$.001$). They trust interest- and location-based groups least ($p$<$.001$).

Why does trust differ by group category?
For friends \& family groups, high trust is a strong sign of bonding social capital~\cite{putnam2000bowling}.
Identity-based groups (e.g., parenting groups) and education \& work groups elicit trust by establishing a shared identity among group members~\cite{moser2017community}.
Finally, interest- and location-based groups may represent less bonding and more bridging social capital~\cite{granovetter1985economic}, especially for information sharing.
People use these groups more as places to transact and exchange (both physical goods and information) than as places to build relationships \cite{granovetter1985economic}.
By comparing groups across different categories, we can develop a more holistic understanding of trust across different types of social groups that also draws on insights from previously isolated studies \cite{moser2017community,holtz2017social}.

\subsubsection{Activity}

Here, we studied both a survey participant's activity in a group as well as the overall group activity across all members.
Measures of activity include time spent in the group and the number of actions (posts, likes, or comments) taken in the group, averaged across the 28 days preceding the survey.
In the case of public groups, activity also included contributions from nonmembers.
An individual's overall site engagement was not predictive of trust, and thus was excluded from our analyses.
Including activity features (time spent, group activity, and participant in group activity) to the baseline model improves its adjusted $R^2$ by $0.04$ ($p$<$.001$).

As many activity features are correlated, we report coefficients when the feature is independently added to the baseline model.
Time spent in the group, both by the individual ($\beta$=$0.04$, $p$<$.001$) and by other group members ($\beta$=$0.05$, $p$<$.001$) independently predicts higher trust in groups.
Overall, the number of posts per member ($\beta$=$0.07$, $p$<$.001$), and the number of likes and comments per post  ($\beta$=$0.07$, $p$<$.001$) were also both independently associated with higher trust.
However, the number of comments and likes a participant made in a group was associated with higher trust ($\beta$=$0.10$ on log scale, $p$<$.001$), but not the number of posts the participant wrote.

Why is this the case?
Posting in a group may be influenced by a variety of factors other than trust (e.g., self-esteem~\cite{forest2012social}).
In contrast, likes and comments are forms of directed communication that people use to maintain relationships with others~\cite{ellison2014cultivating} and may therefore be more conducive to building trust.

\subsubsection{Homogeneity and homophily}
Trust may also be influenced by how similar people in a group are to each other (homogeneity), and how similar an individual is to others in the group (homophily).

For each group, we measured age and gender diversity by computing the gender entropy of the group's members and the standard deviation of their ages.
To measure homophily, we constructed a simple distance measure based on the approach of~\cite{abrahao2017reputation}.
If the participant had the same gender with the majority of the group members, we coded the gender distance as 0, otherwise 1.
If the participant's age was within 5 years of the average age of the group, we coded the age distance as 0, otherwise 1.
The total distance from average group members was calculated as the $L_1$ distance, i.e., the sum of gender and age distance $\in (0, 1, 2)$.
As different types of groups may have different demographic compositions, we controlled for group category in this analysis.

Adding homogeneity and homophily features to the baseline model results in a small improvement (increased adjusted $R^2$ by less than $0.01$, $p$<$.001$).
Nonetheless, we found that both gender ($\beta$=$0.04$, $p$<$.001$)
and age homogeneity ($\beta$=$0.04$, $p$<$.001$) were associated with higher trust.

Surprisingly, homophily, measured as described above, was not predictive of trust in groups. 
This contrasts with findings in previous work on trust and homophily in dyadic exchange, which found that trust increases with gender and age homophily \cite{abrahao2017reputation,ahmad2011trust}.
While we only studied age and gender homophily here, future work may consider other forms of homophily (e.g., with respect to interests, location, or socio-economic status) or other measures of homophily, especially in a group rather than dyadic context.

\subsubsection{Network structure}

To understand how network structure mediates trust, we calculated the following network features for each group:
\begin{enumerate*}
    \item \textit{network density}: the number of friendships in the entire group friendship graph divided by the number of possible combinations;
    \item \textit{average clustering coefficient}:  the average local clustering coefficient in the group membership graph, which measures what proportion of an individual's friends also know one another;
    \item \textit{participant degree centrality}:  the number of friends a participant has in the group, normalized by group size;
    \item \textit{$k$-core existence}: a measure of how a participant's friends in the group are connected with each other, calculated as whether a $k$-core component~\cite{ugander2012structural} exists for participant's friendship graph in the group (we found that $k$=$5$ resulted in the greatest model improvement);
    \item \textit{average mutual friend count}: the mean number of mutual friends between participant and group members. %
\end{enumerate*}

\begin{figure}
    \includegraphics[width=0.90\linewidth]{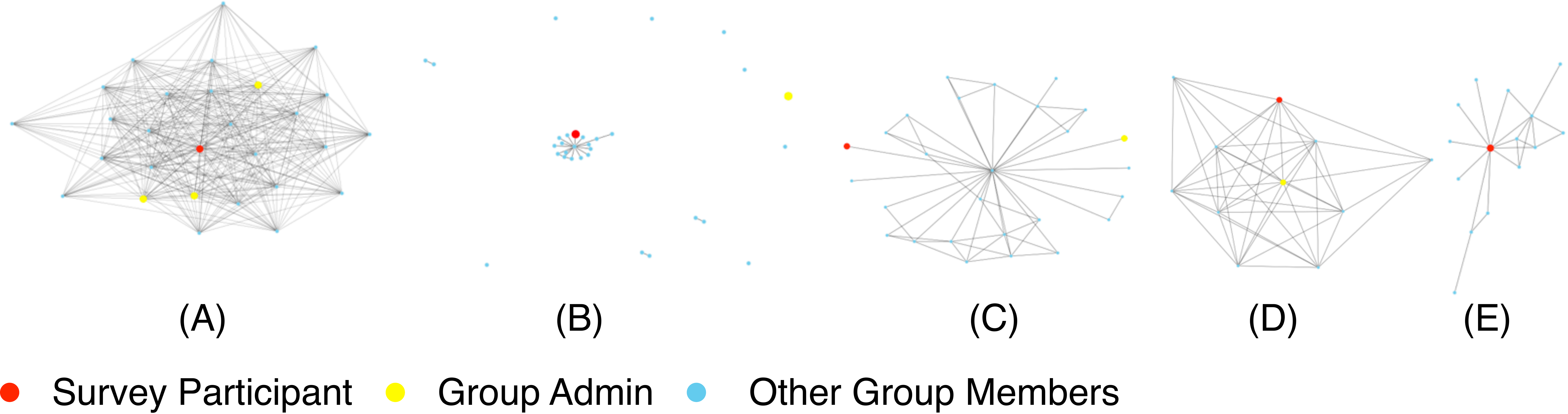}
    \caption{Groups differ in network density, participant degree centrality, and how a participant's friends are linked to each other. Each node represent a group member. Each edge represents a friendship between two members. The survey participant is colored in red, and group admins are colored in yellow.}
    \label{fig:network_structure}
\end{figure}

\autoref{fig:network_structure} illustrates how several group networks in our sample differ along these network features.
For example, Group A has higher network density and higher average clustering coefficient than group B.
Groups C and D differ in the participant's degree centrality. %
Group D contains a $5$-core, but E does not.

These network features, when added to the baseline model, improves its adjusted $R^2$ by $0.10$ ($p$<$.001$).
Each feature was positively associated with trust in groups ($p$<$.001$), though we note that these network features correlate highly with one another.
Considering these features separately, the average clustering coefficient was most predictive ($\beta$=$1.08$, $p$<$.001$), followed by group density ($\beta$=$0.93$, $p$<$.001$) and the participant's degree centrality ($\beta$=$0.84$, $p$<$.001$).

\subsection{Predicting Trust in Groups}

Thus far, we have shown how various sets of group characteristics separately contribute to trust, after controlling for individual characteristics. 
Here, we examine how these features can together predict composite trust in groups.

\begin{figure}[]
\centering 
    \includegraphics[width=0.68\linewidth]{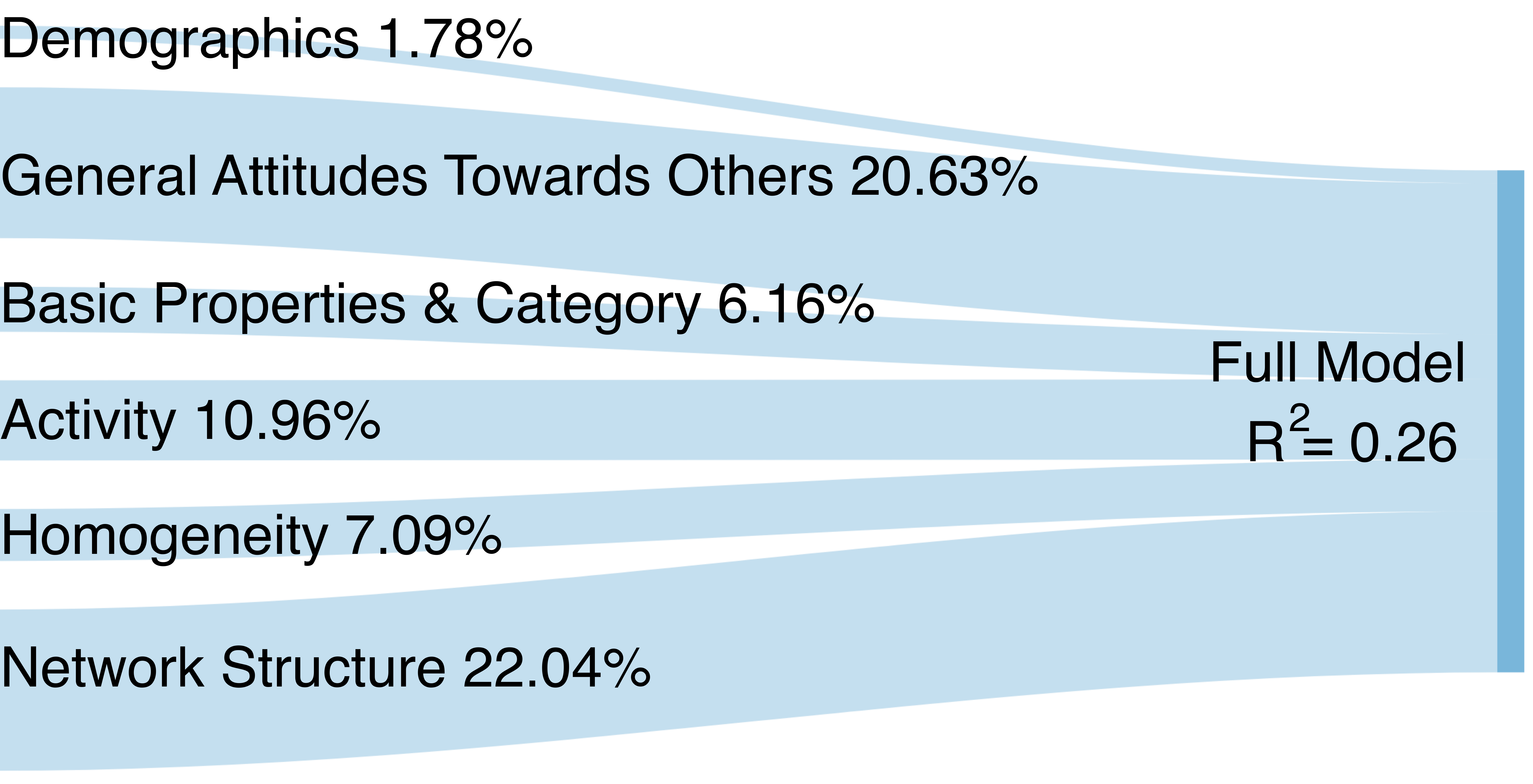}
    \caption{For each feature set, we calculated the average feature importance (measured by relative percent increase in MSE when a feature is removed) in predicting trust in groups. Network structure was the most important, followed by an individual's general attitudes towards others.}
    \label{fig:relative_feature_importance}
\end{figure} 
A random forest model that uses all feature sets (in both~\autoref{tab:baseline} and \autoref{tab:group_features}) reached a performance of out-of-sample adjusted $R^2$ of $0.26$ and a mean-squared error (MSE) of $0.53$. We obtained similar performance using multiple linear regression.

To understand the relative importance of the different feature sets, we ranked all features by how much a random permutation of their values increased the model's MSE.
These values are shown in \autoref{fig:relative_feature_importance}.
We find that network features are most important, followed by an individual's general attitude towards others. Least important were demographic features.
Overall, this result suggests that both individual and group characteristics are important in predicting trust in groups.

\subsubsection{Predicting trust using only observational data}
As we demonstrate relatively robust performance in predicting trust in groups, one might consider using such predictions to make better group or community recommendations.
However, our model uses survey responses about individual differences, including disposition to trust and related concepts, to make predictions about trust in a specific group.
In practical settings, it may not be feasible to administer the survey questions on individual differences to all users.
This motivates the question of how well our modeling approach works in the absence of the individual differences survey features.
Excluding these features, our best model obtained an adjusted $R^2$ of $0.15$ and $MSE$ of $0.59$.
In this model, network structure features were again most important, but instead followed by group activity features.
\subsection{Group Outcomes}

Theoretical accounts of trust emphasize the impact of trust on community outcomes, attributing trust to prosperity~\cite{fukuyama1995trust}, among other things.
Here, we analyze how trust relates to three different group outcomes:
\begin{enumerate*}
    \item the percentage change in group size,
    \item the percentage change in new tie formation (the number of new ties divided by the number of pre-existing ties) among other members of the group, and
    \item the percentage change in new tie formation by the survey participant in the group.
\end{enumerate*}
All three measures were calculated by comparing the state of the group on the day of the survey to that 28 days after.
As these changes tend to be small, with a median change of about 1\%, we instead predict whether each measure would increase by more than 1\%.

\begin{figure}[]
    \includegraphics[width=0.99\linewidth]{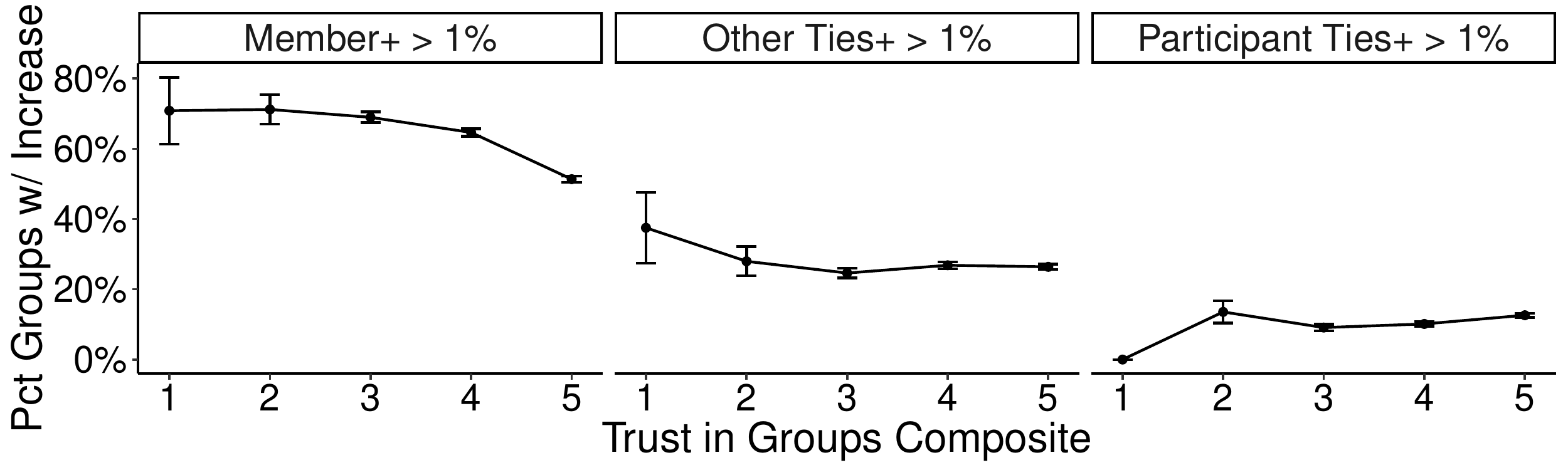}
    \caption{Groups with higher trust ratings are less likely to increase in size (left), more likely for the survey participant to form new connections in them (right), and had no effect on the likelihood on forming friendships among group members other than the rater (center).
    }
    \label{fig:trust_group_outcome}
\end{figure} 
~\autoref{fig:trust_group_outcome} shows the percentage of groups that exhibited an increase by more than 1\% in each of the group outcomes listed above. 
Using logistic regression and controlling for basic group properties such as group size, we found that %
higher trust was associated with a \textit{lower} likelihood of a group increasing in size (odds ratio -0.87, $p$<$.001$); and a \textit{higher} likelihood that the survey participant would form more new friendships in the group (odds ratio 1.29, $p$<$.001$).
Trust in a group was not predictive of the likelihood of \textit{other} group members forming friendships in the group.%

These results suggest a tension between trust and growth for online groups.
Our findings are consistent with previous work on online communities that found that ``cliquishness'' (or high triangle density) makes a community less attractive to join and less likely to grow in size~\cite{backstrom2006group}.
While membership growth is an important indicator of success for online groups~\cite{kraut2014role}, trust, partially elicited by small groups and exclusive membership~\cite{moser2017community}, can limit group expansion (but nonetheless encourages individuals to make new connections within the group).
Future work can examine the relationship between trust and group longevity, as well as other interaction dynamics such as forming sub-communities within the group.

\section{Discussion}
In this paper, we present a framework for predicting an individual's trust in one of their social groups on Facebook.
Combining a large and diverse survey with behavioral and demographic data,
we show that both individual characteristics and group characteristics contribute substantially to trust.
We are able to explain a significant portion of an individual's trust in groups ($R^2$=$0.26$) as well as show how trust relates to outcomes such as membership growth and the formation of new within-group friendships.

This work builds on many previous studies of trust in groups by showing how features previously studied in isolation may interact with each other and how important these features are relative to each other.
Beyond confirming that both an individual's general disposition to trust others \cite{ferguson2015sinking} and a group's size \cite{brewer1991social} affect that individual's trust in a group, we further show that group size matters less to individuals with a greater disposition to trust, and that an individual's feelings of receiving social support from others in general is actually more predictive of trust in groups than their general disposition to trust.
Apart from demonstrating that people do trust smaller, more private groups, we show that among groups with more than 150 members, the effect of exclusive membership decreases.
Where previous work has suggested a relation between group connectivity and trust \cite{coleman1988social,yuki2005cross}, we also demonstrate that network measures such as the average clustering coefficient in a group are among the strongest predictors of trust in a group.
Our findings on how directed communication such as likes and comments contributes to group trust corroborate similar observations in qualitative studies \cite{moser2017community}.

Nonetheless, several null results suggest areas for future exploration.
While prior work suggests trust differs with sociodemographic factors \cite{pew2007americans}, we found that age and gender explain close to zero variance in one's trust in groups.
Future work may consider exploring other factors such as geography or socioeconomic status.
Cultural differences may also play a significant role in trust: prior work found that an indirect relationship between two people was more likely to increase trust for Japanese than Americans~\cite{yuki2005cross}, suggesting that network structure may be more predictive of trust among the former.
Though we found that gender- and age-homogenous groups were more trusted, we also found no evidence that gender or age homophily predicts trust in groups, in contrast to previous literature suggesting that relationships between similar individuals tend to be more trusting \cite{abrahao2017reputation,ahmad2011trust,mcpherson2001birds}.
Understanding the extent to which these findings apply to specific situations --- moms' buy-and-sell groups on Facebook are known to garner trust~\cite{moser2017community} --- remains future work.

Future work may also involve investigating other potential correlates of trust such as psychological safety \cite{edmondson2004psychological} and belonging \cite{zhao2012cultivating}, as well as other outcomes of trust on online groups.
For example, high trust may lead to a greater willingness to attend an event, share (or believe) information originating from within the group, or donate to a cause.

\subsection{Design Implications}
The work reported here has several potential implications for the design of online communities.

We showed that certain types of actions (e.g., commenting and liking) are more positively associated with trust than others (e.g., posting).
This adds nuance to previous findings that people have greater trust in communities in which they are more active ~\cite{cartwright1953group}.
As such, platforms could prioritize facilitating directed interactions among group members, for example, expanding features to support polling, brainstorming, and collective planning.
At the same time, these findings may also inform the design of content recommendation systems.
If these findings indicate that directed communication is a key signal of trust, then incorporating such signals of directed communication may better ensure that people see more content from communities that they trust more.

Consistent with prior work ~\cite{kraut2014role}, we found that trust grows with the number of group admins and decreases with group size.
As online communities grow, it may be beneficial for platforms to encourage groups to recruit additional admins to maintain existing levels of trust.

Further, network properties of online communities such as the average clustering coefficient are strong predictors of trust.
Adding members that increase the average clustering of the group may be beneficial both to new members and to the group as a whole.

Given that trust in a group correlates with behavioral signals, with additional research, platforms may also be able to provide a rough indicator of trust in groups and how it may be changing over time.

Last, our findings suggest alternative strategies for recommending groups to individuals.
For instance, recommending smaller, less popular groups may not only increase the diversity of group recommendations, but also lead to greater trust and user satisfaction.

\subsection{Limitations}
Our analysis is limited to groups on Facebook.
Understanding how trust differs in communities with different policies on anonymity (e.g., Reddit or Nextdoor) or that have different feedback mechanisms remains an important area for future exploration.
Anonymity may increase trust by making it easier for vulnerable populations to talk about sensitive issues, but also have a disinhibiting effect and increase harassment and thus reduce trust \cite{kiesler1984social}; indicators of reputation or popularity such as up-votes and down-votes may also influence trust, especially in the absence of other social signals \cite{resnick2002trust}.
Still, many group properties (e.g., group size) that we examined apply to groups in general; the interactions (e.g., posting or liking) that we looked at are also common on other social media platforms.
Along with the large number and diversity of groups we surveyed, we expect that many of our findings will generalize to other online communities\footnote{Code to reproduce our analysis is available at \url{https://github.com/facebookresearch/trust-in-groups}}.
While we controlled for individual differences such as demographics and an individual's general attitudes towards others, understanding differences that may arise in offline groups and with regards to other factors such as location remains future work. Also, individuals may choose to join groups based on other unobserved differences (e.g., word-of-mouth). Finally, our methodology
is based on correlations between variables and cannot directly suggest causation.
Most significantly, it is possible that individuals who have different propensities to trust tend to join entirely different groups, explaining some of our observed differences. 
Similar limitations apply to the group outcomes analysis.
While greater trust may lead one to connect to other members of a group, it may also arise from making these connections.

\section{Conclusion}
Groups play a significant role in an individual's social experiences and interactions.
Trust, which predicts numerous positive outcomes for a group and its members, is core to a group's proper functioning.
In this paper, we presented a framework for predicting an individual's trust in a social group, and identified characteristics of both the individual and the group that help predict the individual's trust in the group.
This work can contribute to future research and design decisions that better support trust in online communities and foster long-term meaningful interactions online and offline.

\section*{Acknowledgements}
We would like to thank Steve Carter, Nick Brown, Ann Hsieh, Lada Adamic, Moira Burke, Israel Nir, Alex Dow, and our reviewers for their help and feedback.

\balance{}
\bibliographystyle{ACM-Reference-Format}
\bibliography{lib}


\begin{thebibliography}{86}

%%% the Web address.

\ifx \showCODEN    \undefined \def \showCODEN     #1{\unskip}     \fi
\ifx \showDOI      \undefined \def \showDOI       #1{#1}\fi
\ifx \showISBNx    \undefined \def \showISBNx     #1{\unskip}     \fi
\ifx \showISBNxiii \undefined \def \showISBNxiii  #1{\unskip}     \fi
\ifx \showISSN     \undefined \def \showISSN      #1{\unskip}     \fi
\ifx \showLCCN     \undefined \def \showLCCN      #1{\unskip}     \fi
\ifx \shownote     \undefined \def \shownote      #1{#1}          \fi
\ifx \showarticletitle \undefined \def \showarticletitle #1{#1}   \fi
\ifx \showURL      \undefined \def \showURL       {\relax}        \fi
% invisible to TeX
\providecommand\bibfield[2]{#2}
\providecommand\bibinfo[2]{#2}
\providecommand\natexlab[1]{#1}
\providecommand\showeprint[2][]{arXiv:#2}

\bibitem[\protect\citeauthoryear{Abrahao, Parigi, Gupta, and Cook}{Abrahao
  et~al\mbox{.}}{2017}]%
        {abrahao2017reputation}
\bibfield{author}{\bibinfo{person}{Bruno Abrahao}, \bibinfo{person}{Paolo
  Parigi}, \bibinfo{person}{Alok Gupta}, {and} \bibinfo{person}{Karen~S Cook}.}
  \bibinfo{year}{2017}\natexlab{}.
\newblock \showarticletitle{Reputation offsets trust judgments based on social
  biases among Airbnb users}.
\newblock \bibinfo{journal}{\emph{Proceedings of the National Academy of
  Sciences}} \bibinfo{volume}{114}, \bibinfo{number}{37}
  (\bibinfo{year}{2017}), \bibinfo{pages}{9848--9853}.
\newblock


\bibitem[\protect\citeauthoryear{Ahmad, Ahmed, Srivastava, and Poole}{Ahmad
  et~al\mbox{.}}{2011}]%
        {ahmad2011trust}
\bibfield{author}{\bibinfo{person}{Muhammad~A Ahmad}, \bibinfo{person}{Iftekhar
  Ahmed}, \bibinfo{person}{Jaideep Srivastava}, {and}
  \bibinfo{person}{Marshall~S Poole}.} \bibinfo{year}{2011}\natexlab{}.
\newblock \showarticletitle{Trust me, I'm an expert: Trust, homophily and
  expertise in MMOs}. In \bibinfo{booktitle}{\emph{Proceedings of the IEEE
  International Conference on Privacy, Security, Risk and Trust}}. IEEE,
  \bibinfo{pages}{882--887}.
\newblock


\bibitem[\protect\citeauthoryear{Ainsworth, Blehar, Waters, and Wall}{Ainsworth
  et~al\mbox{.}}{2015}]%
        {ainsworth2015patterns}
\bibfield{author}{\bibinfo{person}{Mary~DS Ainsworth}, \bibinfo{person}{Mary~C
  Blehar}, \bibinfo{person}{Everett Waters}, {and} \bibinfo{person}{Sally~N
  Wall}.} \bibinfo{year}{2015}\natexlab{}.
\newblock \bibinfo{booktitle}{\emph{Patterns of attachment: A psychological
  study of the strange situation}}.
\newblock \bibinfo{publisher}{Psychology Press}.
\newblock


\bibitem[\protect\citeauthoryear{Ashleigh, Higgs, and Dulewicz}{Ashleigh
  et~al\mbox{.}}{2012}]%
        {ashleigh2012new}
\bibfield{author}{\bibinfo{person}{Melanie~J Ashleigh},
  \bibinfo{person}{Malcolm Higgs}, {and} \bibinfo{person}{Vic Dulewicz}.}
  \bibinfo{year}{2012}\natexlab{}.
\newblock \showarticletitle{A new propensity to trust scale and its
  relationship with individual well-being: implications for HRM policies and
  practices}.
\newblock \bibinfo{journal}{\emph{Human Resource Management Journal}}
  \bibinfo{volume}{22}, \bibinfo{number}{4} (\bibinfo{year}{2012}),
  \bibinfo{pages}{360--376}.
\newblock


\bibitem[\protect\citeauthoryear{Bachmann}{Bachmann}{2001}]%
        {bachmann2001trust}
\bibfield{author}{\bibinfo{person}{Reinhard Bachmann}.}
  \bibinfo{year}{2001}\natexlab{}.
\newblock \showarticletitle{Trust, power and control in trans-organizational
  relations}.
\newblock \bibinfo{journal}{\emph{Organization Studies}} \bibinfo{volume}{22},
  \bibinfo{number}{2} (\bibinfo{year}{2001}), \bibinfo{pages}{337--365}.
\newblock


\bibitem[\protect\citeauthoryear{Backstrom, Huttenlocher, Kleinberg, and
  Lan}{Backstrom et~al\mbox{.}}{2006}]%
        {backstrom2006group}
\bibfield{author}{\bibinfo{person}{Lars Backstrom}, \bibinfo{person}{Dan
  Huttenlocher}, \bibinfo{person}{Jon Kleinberg}, {and}
  \bibinfo{person}{Xiangyang Lan}.} \bibinfo{year}{2006}\natexlab{}.
\newblock \showarticletitle{Group formation in large social networks:
  membership, growth, and evolution}. In \bibinfo{booktitle}{\emph{Proceedings
  of the ACM International Conference on Knowledge Discovery and Data Mining}}.
  ACM, \bibinfo{pages}{44--54}.
\newblock


\bibitem[\protect\citeauthoryear{Barrera~Jr and Ainlay}{Barrera~Jr and
  Ainlay}{1983}]%
        {barrera1983structure}
\bibfield{author}{\bibinfo{person}{Manuel Barrera~Jr} {and}
  \bibinfo{person}{Sheila~L Ainlay}.} \bibinfo{year}{1983}\natexlab{}.
\newblock \showarticletitle{The structure of social support: A conceptual and
  empirical analysis}.
\newblock \bibinfo{journal}{\emph{Journal of Community Psychology}}
  \bibinfo{volume}{11}, \bibinfo{number}{2} (\bibinfo{year}{1983}),
  \bibinfo{pages}{133--143}.
\newblock


\bibitem[\protect\citeauthoryear{Berg, Dickhaut, and McCabe}{Berg
  et~al\mbox{.}}{1995}]%
        {berg1995trust}
\bibfield{author}{\bibinfo{person}{Joyce Berg}, \bibinfo{person}{John
  Dickhaut}, {and} \bibinfo{person}{Kevin McCabe}.}
  \bibinfo{year}{1995}\natexlab{}.
\newblock \showarticletitle{Trust, reciprocity, and social history}.
\newblock \bibinfo{journal}{\emph{Games and Economic Behavior}}
  \bibinfo{volume}{10}, \bibinfo{number}{1} (\bibinfo{year}{1995}),
  \bibinfo{pages}{122--142}.
\newblock


\bibitem[\protect\citeauthoryear{Bj{\o}rnskov}{Bj{\o}rnskov}{2007}]%
        {bjornskov2007determinants}
\bibfield{author}{\bibinfo{person}{Christian Bj{\o}rnskov}.}
  \bibinfo{year}{2007}\natexlab{}.
\newblock \showarticletitle{Determinants of generalized trust: A cross-country
  comparison}.
\newblock \bibinfo{journal}{\emph{Public Choice}} \bibinfo{volume}{130},
  \bibinfo{number}{1-2} (\bibinfo{year}{2007}), \bibinfo{pages}{1--21}.
\newblock


\bibitem[\protect\citeauthoryear{Blau}{Blau}{1964}]%
        {blau1964power}
\bibfield{author}{\bibinfo{person}{Peter Blau}.}
  \bibinfo{year}{1964}\natexlab{}.
\newblock \bibinfo{booktitle}{\emph{Exchange and power in social life}}.
\newblock \bibinfo{publisher}{John Wiley \& Sons}.
\newblock


\bibitem[\protect\citeauthoryear{Boss}{Boss}{1978}]%
        {boss1978trust}
\bibfield{author}{\bibinfo{person}{R~Wayne Boss}.}
  \bibinfo{year}{1978}\natexlab{}.
\newblock \showarticletitle{Trust and managerial problem solving revisited}.
\newblock \bibinfo{journal}{\emph{Group \& Organization Studies}}
  \bibinfo{volume}{3}, \bibinfo{number}{3} (\bibinfo{year}{1978}),
  \bibinfo{pages}{331--342}.
\newblock


\bibitem[\protect\citeauthoryear{Bowlby}{Bowlby}{1969}]%
        {bowlby1969attachment}
\bibfield{author}{\bibinfo{person}{John Bowlby}.}
  \bibinfo{year}{1969}\natexlab{}.
\newblock \bibinfo{booktitle}{\emph{Attachment and Loss: Attachment}}.
\newblock \bibinfo{publisher}{Basic books}.
\newblock


\bibitem[\protect\citeauthoryear{Brewer}{Brewer}{1991}]%
        {brewer1991social}
\bibfield{author}{\bibinfo{person}{Marilynn~B Brewer}.}
  \bibinfo{year}{1991}\natexlab{}.
\newblock \showarticletitle{The social self: On being the same and different at
  the same time}.
\newblock \bibinfo{journal}{\emph{Personality and Social psychology Bulletin}}
  \bibinfo{volume}{17}, \bibinfo{number}{5} (\bibinfo{year}{1991}),
  \bibinfo{pages}{475--482}.
\newblock


\bibitem[\protect\citeauthoryear{Butler~Jr}{Butler~Jr}{1999}]%
        {butler1999trust}
\bibfield{author}{\bibinfo{person}{John~K Butler~Jr}.}
  \bibinfo{year}{1999}\natexlab{}.
\newblock \showarticletitle{Trust expectations, information sharing, climate of
  trust, and negotiation effectiveness and efficiency}.
\newblock \bibinfo{journal}{\emph{Group \& Organization Management}}
  \bibinfo{volume}{24}, \bibinfo{number}{2} (\bibinfo{year}{1999}),
  \bibinfo{pages}{217--238}.
\newblock


\bibitem[\protect\citeauthoryear{Cartwright and Zander}{Cartwright and
  Zander}{1953}]%
        {cartwright1953group}
\bibfield{author}{\bibinfo{person}{Dorwin Cartwright} {and}
  \bibinfo{person}{Alvin Zander}.} \bibinfo{year}{1953}\natexlab{}.
\newblock \showarticletitle{Group cohesiveness: introduction}.
\newblock \bibinfo{journal}{\emph{Group Dynamics: Research and Theory.
  Evanston, IL: Row Peterson}} (\bibinfo{year}{1953}).
\newblock


\bibitem[\protect\citeauthoryear{Center}{Center}{2007}]%
        {pew2007americans}
\bibfield{author}{\bibinfo{person}{Pew~Research Center}.}
  \bibinfo{year}{2007}\natexlab{}.
\newblock \bibinfo{title}{Americans and Social Trust: Who, Where and Why | Pew
  Research Center}.
\newblock
  \bibinfo{howpublished}{\url{http://www.pewsocialtrends.org/2007/02/22/americans-and-social-trust-who-where-and-why}}.
\newblock
\newblock
\shownote{(Accessed Sep 2018).}


\bibitem[\protect\citeauthoryear{Coleman}{Coleman}{1988}]%
        {coleman1988social}
\bibfield{author}{\bibinfo{person}{James~S Coleman}.}
  \bibinfo{year}{1988}\natexlab{}.
\newblock \showarticletitle{Social capital in the creation of human capital}.
\newblock \bibinfo{journal}{\emph{Amer. J. Sociology}}  \bibinfo{volume}{94}
  (\bibinfo{year}{1988}), \bibinfo{pages}{S95--S120}.
\newblock


\bibitem[\protect\citeauthoryear{Colquitt, Scott, and LePine}{Colquitt
  et~al\mbox{.}}{2007}]%
        {colquitt2007trust}
\bibfield{author}{\bibinfo{person}{Jason~A Colquitt}, \bibinfo{person}{Brent~A
  Scott}, {and} \bibinfo{person}{Jeffery~A LePine}.}
  \bibinfo{year}{2007}\natexlab{}.
\newblock \showarticletitle{Trust, trustworthiness, and trust propensity: a
  meta-analytic test of their unique relationships with risk taking and job
  performance.}
\newblock \bibinfo{journal}{\emph{Journal of Applied Psychology}}
  \bibinfo{volume}{92}, \bibinfo{number}{4} (\bibinfo{year}{2007}),
  \bibinfo{pages}{909}.
\newblock


\bibitem[\protect\citeauthoryear{Cook, Yamagishi, Cheshire, Cooper, Matsuda,
  and Mashima}{Cook et~al\mbox{.}}{2005}]%
        {cook2005trust}
\bibfield{author}{\bibinfo{person}{Karen~S Cook}, \bibinfo{person}{Toshio
  Yamagishi}, \bibinfo{person}{Coye Cheshire}, \bibinfo{person}{Robin Cooper},
  \bibinfo{person}{Masafumi Matsuda}, {and} \bibinfo{person}{Rie Mashima}.}
  \bibinfo{year}{2005}\natexlab{}.
\newblock \showarticletitle{Trust building via risk taking: A cross-societal
  experiment}.
\newblock \bibinfo{journal}{\emph{Social Psychology Quarterly}}
  \bibinfo{volume}{68}, \bibinfo{number}{2} (\bibinfo{year}{2005}),
  \bibinfo{pages}{121--142}.
\newblock


\bibitem[\protect\citeauthoryear{Denson, Lickel, Curtis, Stenstrom, and
  Ames}{Denson et~al\mbox{.}}{2006}]%
        {denson2006roles}
\bibfield{author}{\bibinfo{person}{Thomas~F Denson}, \bibinfo{person}{Brian
  Lickel}, \bibinfo{person}{Mathew Curtis}, \bibinfo{person}{Douglas~M
  Stenstrom}, {and} \bibinfo{person}{Daniel~R Ames}.}
  \bibinfo{year}{2006}\natexlab{}.
\newblock \showarticletitle{The roles of entitativity and essentiality in
  judgments of collective responsibility}.
\newblock \bibinfo{journal}{\emph{Group Processes \& Intergroup Relations}}
  \bibinfo{volume}{9}, \bibinfo{number}{1} (\bibinfo{year}{2006}),
  \bibinfo{pages}{43--61}.
\newblock


\bibitem[\protect\citeauthoryear{Denters}{Denters}{2002}]%
        {denters2002size}
\bibfield{author}{\bibinfo{person}{Bas Denters}.}
  \bibinfo{year}{2002}\natexlab{}.
\newblock \showarticletitle{Size and political trust: evidence from Denmark,
  the Netherlands, Norway, and the United Kingdom}.
\newblock \bibinfo{journal}{\emph{Environment and Planning C: Government and
  Policy}} \bibinfo{volume}{20}, \bibinfo{number}{6} (\bibinfo{year}{2002}),
  \bibinfo{pages}{793--812}.
\newblock


\bibitem[\protect\citeauthoryear{Dirks}{Dirks}{1999}]%
        {dirks1999effects}
\bibfield{author}{\bibinfo{person}{Kurt~T Dirks}.}
  \bibinfo{year}{1999}\natexlab{}.
\newblock \showarticletitle{The effects of interpersonal trust on work group
  performance.}
\newblock \bibinfo{journal}{\emph{Journal of Applied Psychology}}
  \bibinfo{volume}{84}, \bibinfo{number}{3} (\bibinfo{year}{1999}),
  \bibinfo{pages}{445}.
\newblock


\bibitem[\protect\citeauthoryear{Dunbar}{Dunbar}{1992}]%
        {dunbar1992neocortex}
\bibfield{author}{\bibinfo{person}{Robin~IM Dunbar}.}
  \bibinfo{year}{1992}\natexlab{}.
\newblock \showarticletitle{Neocortex size as a constraint on group size in
  primates}.
\newblock \bibinfo{journal}{\emph{Journal of Human Evolution}}
  \bibinfo{volume}{22}, \bibinfo{number}{6} (\bibinfo{year}{1992}),
  \bibinfo{pages}{469--493}.
\newblock


\bibitem[\protect\citeauthoryear{Edmondson, Kramer, and Cook}{Edmondson
  et~al\mbox{.}}{2004}]%
        {edmondson2004psychological}
\bibfield{author}{\bibinfo{person}{Amy~C Edmondson},
  \bibinfo{person}{Roderick~M Kramer}, {and} \bibinfo{person}{Karen~S Cook}.}
  \bibinfo{year}{2004}\natexlab{}.
\newblock \showarticletitle{Psychological safety, trust, and learning in
  organizations: A group-level lens}.
\newblock \bibinfo{journal}{\emph{Trust and Distrust in Organizations: Dilemmas
  and Approaches}}  \bibinfo{volume}{12} (\bibinfo{year}{2004}),
  \bibinfo{pages}{239--272}.
\newblock


\bibitem[\protect\citeauthoryear{Ellison, Vitak, Gray, and Lampe}{Ellison
  et~al\mbox{.}}{2014}]%
        {ellison2014cultivating}
\bibfield{author}{\bibinfo{person}{Nicole~B Ellison}, \bibinfo{person}{Jessica
  Vitak}, \bibinfo{person}{Rebecca Gray}, {and} \bibinfo{person}{Cliff Lampe}.}
  \bibinfo{year}{2014}\natexlab{}.
\newblock \showarticletitle{Cultivating social resources on social network
  sites: Facebook relationship maintenance behaviors and their role in social
  capital processes}.
\newblock \bibinfo{journal}{\emph{Journal of Computer-Mediated Communication}}
  \bibinfo{volume}{19}, \bibinfo{number}{4} (\bibinfo{year}{2014}),
  \bibinfo{pages}{855--870}.
\newblock


\bibitem[\protect\citeauthoryear{Emerson}{Emerson}{1976}]%
        {emerson1976social}
\bibfield{author}{\bibinfo{person}{Richard~M Emerson}.}
  \bibinfo{year}{1976}\natexlab{}.
\newblock \showarticletitle{Social exchange theory}.
\newblock \bibinfo{journal}{\emph{Annual Review of Sociology}}
  \bibinfo{volume}{2}, \bibinfo{number}{1} (\bibinfo{year}{1976}),
  \bibinfo{pages}{335--362}.
\newblock


\bibitem[\protect\citeauthoryear{Facebook}{Facebook}{2018}]%
        {facebook2018help}
\bibfield{author}{\bibinfo{person}{Facebook}.} \bibinfo{year}{2018}\natexlab{}.
\newblock \bibinfo{title}{Facebook Help Center}.
\newblock
  \bibinfo{howpublished}{\url{https://www.facebook.com/help/1629740080681586}}.
\newblock
\newblock
\shownote{(Accessed Sep 2018).}


\bibitem[\protect\citeauthoryear{Ferguson and Peterson}{Ferguson and
  Peterson}{2015}]%
        {ferguson2015sinking}
\bibfield{author}{\bibinfo{person}{Amanda~J Ferguson} {and}
  \bibinfo{person}{Randall~S Peterson}.} \bibinfo{year}{2015}\natexlab{}.
\newblock \showarticletitle{Sinking slowly: Diversity in propensity to trust
  predicts downward trust spirals in small groups.}
\newblock \bibinfo{journal}{\emph{Journal of Applied Psychology}}
  \bibinfo{volume}{100}, \bibinfo{number}{4} (\bibinfo{year}{2015}),
  \bibinfo{pages}{1012}.
\newblock


\bibitem[\protect\citeauthoryear{Fine and Holyfield}{Fine and
  Holyfield}{1996}]%
        {fine1996secrecy}
\bibfield{author}{\bibinfo{person}{Gary~A Fine} {and} \bibinfo{person}{Lori
  Holyfield}.} \bibinfo{year}{1996}\natexlab{}.
\newblock \showarticletitle{Secrecy, trust, and dangerous leisure: Generating
  group cohesion in voluntary organizations}.
\newblock \bibinfo{journal}{\emph{Social Psychology Quarterly}}
  (\bibinfo{year}{1996}), \bibinfo{pages}{22--38}.
\newblock


\bibitem[\protect\citeauthoryear{Forest and Wood}{Forest and Wood}{2012}]%
        {forest2012social}
\bibfield{author}{\bibinfo{person}{Amanda~L Forest} {and}
  \bibinfo{person}{Joanne~V Wood}.} \bibinfo{year}{2012}\natexlab{}.
\newblock \showarticletitle{When social networking is not working: Individuals
  with low self-esteem recognize but do not reap the benefits of
  self-disclosure on Facebook}.
\newblock \bibinfo{journal}{\emph{Psychological Science}} \bibinfo{volume}{23},
  \bibinfo{number}{3} (\bibinfo{year}{2012}), \bibinfo{pages}{295--302}.
\newblock


\bibitem[\protect\citeauthoryear{Fukuyama}{Fukuyama}{1995}]%
        {fukuyama1995trust}
\bibfield{author}{\bibinfo{person}{Francis Fukuyama}.}
  \bibinfo{year}{1995}\natexlab{}.
\newblock \bibinfo{booktitle}{\emph{Trust: The social virtues and the creation
  of prosperity}}.
\newblock \bibinfo{publisher}{Free Press}.
\newblock


\bibitem[\protect\citeauthoryear{Gambetta}{Gambetta}{1988}]%
        {gambetta1988trust}
\bibfield{author}{\bibinfo{person}{Diego Gambetta}.}
  \bibinfo{year}{1988}\natexlab{}.
\newblock \bibinfo{booktitle}{\emph{Trust: Making and breaking cooperative
  relations}}.
\newblock \bibinfo{publisher}{Blackwell Publishing Limited}.
\newblock


\bibitem[\protect\citeauthoryear{Gill, Boies, Finegan, and McNally}{Gill
  et~al\mbox{.}}{2005}]%
        {gill2005antecedents}
\bibfield{author}{\bibinfo{person}{Harjinder Gill}, \bibinfo{person}{Kathleen
  Boies}, \bibinfo{person}{Joan~E Finegan}, {and} \bibinfo{person}{Jeffrey
  McNally}.} \bibinfo{year}{2005}\natexlab{}.
\newblock \showarticletitle{Antecedents of trust: Establishing a boundary
  condition for the relation between propensity to trust and intention to
  trust}.
\newblock \bibinfo{journal}{\emph{Journal of Business and Psychology}}
  \bibinfo{volume}{19}, \bibinfo{number}{3} (\bibinfo{year}{2005}),
  \bibinfo{pages}{287--302}.
\newblock


\bibitem[\protect\citeauthoryear{Granovetter}{Granovetter}{1985}]%
        {granovetter1985economic}
\bibfield{author}{\bibinfo{person}{Mark Granovetter}.}
  \bibinfo{year}{1985}\natexlab{}.
\newblock \showarticletitle{Economic action and social structure: The problem
  of embeddedness}.
\newblock \bibinfo{journal}{\emph{Amer. J. Sociology}} \bibinfo{volume}{91},
  \bibinfo{number}{3} (\bibinfo{year}{1985}), \bibinfo{pages}{481--510}.
\newblock


\bibitem[\protect\citeauthoryear{Gulati}{Gulati}{1995}]%
        {gulati1995does}
\bibfield{author}{\bibinfo{person}{Ranjay Gulati}.}
  \bibinfo{year}{1995}\natexlab{}.
\newblock \showarticletitle{Does familiarity breed trust? The implications of
  repeated ties for contractual choice in alliances}.
\newblock \bibinfo{journal}{\emph{Academy of Management Journal}}
  \bibinfo{volume}{38}, \bibinfo{number}{1} (\bibinfo{year}{1995}),
  \bibinfo{pages}{85--112}.
\newblock


\bibitem[\protect\citeauthoryear{Hether, Murphy, and Valente}{Hether
  et~al\mbox{.}}{2014}]%
        {hether2014s}
\bibfield{author}{\bibinfo{person}{Heather~J Hether}, \bibinfo{person}{Sheila~T
  Murphy}, {and} \bibinfo{person}{Thomas~W Valente}.}
  \bibinfo{year}{2014}\natexlab{}.
\newblock \showarticletitle{It's better to give than to receive: The role of
  social support, trust, and participation on health-related social networking
  sites}.
\newblock \bibinfo{journal}{\emph{Journal of Health Communication}}
  \bibinfo{volume}{19}, \bibinfo{number}{12} (\bibinfo{year}{2014}),
  \bibinfo{pages}{1424--1439}.
\newblock


\bibitem[\protect\citeauthoryear{Hogg}{Hogg}{1993}]%
        {hogg1993group}
\bibfield{author}{\bibinfo{person}{Michael~A Hogg}.}
  \bibinfo{year}{1993}\natexlab{}.
\newblock \showarticletitle{Group cohesiveness: A critical review and some new
  directions}.
\newblock \bibinfo{journal}{\emph{European Review of Social Psychology}}
  \bibinfo{volume}{4}, \bibinfo{number}{1} (\bibinfo{year}{1993}),
  \bibinfo{pages}{85--111}.
\newblock


\bibitem[\protect\citeauthoryear{Holtz, MacLean, and Aral}{Holtz
  et~al\mbox{.}}{2017}]%
        {holtz2017social}
\bibfield{author}{\bibinfo{person}{David Holtz}, \bibinfo{person}{Diana~L
  MacLean}, {and} \bibinfo{person}{Sinan Aral}.}
  \bibinfo{year}{2017}\natexlab{}.
\newblock \showarticletitle{Social structure and trust in massive digital
  markets}. In \bibinfo{booktitle}{\emph{Proceedings of the International
  Conference on Information Systems}}.
\newblock


\bibitem[\protect\citeauthoryear{Homans}{Homans}{1958}]%
        {homans1958social}
\bibfield{author}{\bibinfo{person}{George~C Homans}.}
  \bibinfo{year}{1958}\natexlab{}.
\newblock \showarticletitle{Social behavior as exchange}.
\newblock \bibinfo{journal}{\emph{Amer. J. Sociology}} \bibinfo{volume}{63},
  \bibinfo{number}{6} (\bibinfo{year}{1958}), \bibinfo{pages}{597--606}.
\newblock


\bibitem[\protect\citeauthoryear{Jarvenpaa and Leidner}{Jarvenpaa and
  Leidner}{1999}]%
        {jarvenpaa1999communication}
\bibfield{author}{\bibinfo{person}{Sirkka~L Jarvenpaa} {and}
  \bibinfo{person}{Dorothy~E Leidner}.} \bibinfo{year}{1999}\natexlab{}.
\newblock \showarticletitle{Communication and trust in global virtual teams}.
\newblock \bibinfo{journal}{\emph{Organization Science}} \bibinfo{volume}{10},
  \bibinfo{number}{6} (\bibinfo{year}{1999}), \bibinfo{pages}{791--815}.
\newblock


\bibitem[\protect\citeauthoryear{Johnson-George and Swap}{Johnson-George and
  Swap}{1982}]%
        {johnson1982measurement}
\bibfield{author}{\bibinfo{person}{Cynthia Johnson-George} {and}
  \bibinfo{person}{Walter~C Swap}.} \bibinfo{year}{1982}\natexlab{}.
\newblock \showarticletitle{Measurement of specific interpersonal trust:
  Construction and validation of a scale to assess trust in a specific other.}
\newblock \bibinfo{journal}{\emph{Journal of Personality and Social
  Psychology}} \bibinfo{volume}{43}, \bibinfo{number}{6}
  (\bibinfo{year}{1982}), \bibinfo{pages}{1306}.
\newblock


\bibitem[\protect\citeauthoryear{Jones and Leonard}{Jones and Leonard}{2008}]%
        {jones2008trust}
\bibfield{author}{\bibinfo{person}{Kiku Jones} {and} \bibinfo{person}{Lori~NK
  Leonard}.} \bibinfo{year}{2008}\natexlab{}.
\newblock \showarticletitle{Trust in consumer-to-consumer electronic commerce}.
\newblock \bibinfo{journal}{\emph{Information \& Management}}
  \bibinfo{volume}{45}, \bibinfo{number}{2} (\bibinfo{year}{2008}),
  \bibinfo{pages}{88--95}.
\newblock


\bibitem[\protect\citeauthoryear{Kantsperger and Kunz}{Kantsperger and
  Kunz}{2010}]%
        {kantsperger2010consumer}
\bibfield{author}{\bibinfo{person}{Roland Kantsperger} {and}
  \bibinfo{person}{Werner~H Kunz}.} \bibinfo{year}{2010}\natexlab{}.
\newblock \showarticletitle{Consumer trust in service companies: a multiple
  mediating analysis}.
\newblock \bibinfo{journal}{\emph{Managing Service Quality: An International
  Journal}} \bibinfo{volume}{20}, \bibinfo{number}{1} (\bibinfo{year}{2010}),
  \bibinfo{pages}{4--25}.
\newblock


\bibitem[\protect\citeauthoryear{Kiesler, Siegel, and McGuire}{Kiesler
  et~al\mbox{.}}{1984}]%
        {kiesler1984social}
\bibfield{author}{\bibinfo{person}{Sara Kiesler}, \bibinfo{person}{Jane
  Siegel}, {and} \bibinfo{person}{Timothy~W McGuire}.}
  \bibinfo{year}{1984}\natexlab{}.
\newblock \showarticletitle{Social psychological aspects of computer-mediated
  communication.}
\newblock \bibinfo{journal}{\emph{American Psychologist}} \bibinfo{volume}{39},
  \bibinfo{number}{10} (\bibinfo{year}{1984}), \bibinfo{pages}{1123}.
\newblock


\bibitem[\protect\citeauthoryear{Kraut and Fiore}{Kraut and Fiore}{2014}]%
        {kraut2014role}
\bibfield{author}{\bibinfo{person}{Robert~E Kraut} {and}
  \bibinfo{person}{Andrew~T Fiore}.} \bibinfo{year}{2014}\natexlab{}.
\newblock \showarticletitle{The role of founders in building online groups}. In
  \bibinfo{booktitle}{\emph{Proceedings of the ACM Conference on Computer
  Supported Cooperative Work \& Social Computing}}. ACM,
  \bibinfo{pages}{722--732}.
\newblock


\bibitem[\protect\citeauthoryear{Kuwabara}{Kuwabara}{2015}]%
        {kuwabara2015reputation}
\bibfield{author}{\bibinfo{person}{Ko Kuwabara}.}
  \bibinfo{year}{2015}\natexlab{}.
\newblock \showarticletitle{Do reputation systems undermine trust? Divergent
  effects of enforcement type on generalized trust and trustworthiness}.
\newblock \bibinfo{journal}{\emph{Amer. J. Sociology}} \bibinfo{volume}{120},
  \bibinfo{number}{5} (\bibinfo{year}{2015}), \bibinfo{pages}{1390--1428}.
\newblock


\bibitem[\protect\citeauthoryear{La~Macchia, Louis, Hornsey, and
  Leonardelli}{La~Macchia et~al\mbox{.}}{2016}]%
        {la2016small}
\bibfield{author}{\bibinfo{person}{Stephen~T La~Macchia},
  \bibinfo{person}{Winnifred~R Louis}, \bibinfo{person}{Matthew~J Hornsey},
  {and} \bibinfo{person}{Geoffrey~J Leonardelli}.}
  \bibinfo{year}{2016}\natexlab{}.
\newblock \showarticletitle{In small we trust: Lay theories about small and
  large groups}.
\newblock \bibinfo{journal}{\emph{Personality and Social Psychology Bulletin}}
  \bibinfo{volume}{42}, \bibinfo{number}{10} (\bibinfo{year}{2016}),
  \bibinfo{pages}{1321--1334}.
\newblock


\bibitem[\protect\citeauthoryear{Ma, Hancock, Lim~Mingjie, and Naaman}{Ma
  et~al\mbox{.}}{2017}]%
        {ma2017self}
\bibfield{author}{\bibinfo{person}{Xiao Ma}, \bibinfo{person}{Jeffery~T
  Hancock}, \bibinfo{person}{Kenneth Lim~Mingjie}, {and} \bibinfo{person}{Mor
  Naaman}.} \bibinfo{year}{2017}\natexlab{}.
\newblock \showarticletitle{Self-disclosure and perceived trustworthiness of
  Airbnb host profiles}. In \bibinfo{booktitle}{\emph{Proceedings of the ACM
  Conference on Computer Supported Cooperative Work \& Social Computing}}.
  \bibinfo{publisher}{ACM}, \bibinfo{pages}{2397--2409}.
\newblock


\bibitem[\protect\citeauthoryear{Mayer, Davis, and Schoorman}{Mayer
  et~al\mbox{.}}{1995}]%
        {mayer1995integrative}
\bibfield{author}{\bibinfo{person}{Roger~C Mayer}, \bibinfo{person}{James~H
  Davis}, {and} \bibinfo{person}{F~David Schoorman}.}
  \bibinfo{year}{1995}\natexlab{}.
\newblock \showarticletitle{An integrative model of organizational trust}.
\newblock \bibinfo{journal}{\emph{Academy of Management Review}}
  \bibinfo{volume}{20}, \bibinfo{number}{3} (\bibinfo{year}{1995}),
  \bibinfo{pages}{709--734}.
\newblock


\bibitem[\protect\citeauthoryear{McEvily, Perrone, and Zaheer}{McEvily
  et~al\mbox{.}}{2003}]%
        {mcevily2003trust}
\bibfield{author}{\bibinfo{person}{Bill McEvily}, \bibinfo{person}{Vincenzo
  Perrone}, {and} \bibinfo{person}{Akbar Zaheer}.}
  \bibinfo{year}{2003}\natexlab{}.
\newblock \showarticletitle{Trust as an organizing principle}.
\newblock \bibinfo{journal}{\emph{Organization Science}} \bibinfo{volume}{14},
  \bibinfo{number}{1} (\bibinfo{year}{2003}), \bibinfo{pages}{91--103}.
\newblock


\bibitem[\protect\citeauthoryear{McPherson, Smith-Lovin, and Cook}{McPherson
  et~al\mbox{.}}{2001}]%
        {mcpherson2001birds}
\bibfield{author}{\bibinfo{person}{Miller McPherson}, \bibinfo{person}{Lynn
  Smith-Lovin}, {and} \bibinfo{person}{James~M Cook}.}
  \bibinfo{year}{2001}\natexlab{}.
\newblock \showarticletitle{Birds of a feather: Homophily in social networks}.
\newblock \bibinfo{journal}{\emph{Annual Review of Sociology}}
  \bibinfo{volume}{27}, \bibinfo{number}{1} (\bibinfo{year}{2001}),
  \bibinfo{pages}{415--444}.
\newblock


\bibitem[\protect\citeauthoryear{Meyerson, Weick, and Kramer}{Meyerson
  et~al\mbox{.}}{1996}]%
        {meyerson1996swift}
\bibfield{author}{\bibinfo{person}{Debra Meyerson}, \bibinfo{person}{Karl~E
  Weick}, {and} \bibinfo{person}{Roderick~M Kramer}.}
  \bibinfo{year}{1996}\natexlab{}.
\newblock \showarticletitle{Swift trust and temporary groups}.
\newblock \bibinfo{journal}{\emph{Trust in Organizations: Frontiers of Theory
  and Research}}  \bibinfo{volume}{166} (\bibinfo{year}{1996}),
  \bibinfo{pages}{195}.
\newblock


\bibitem[\protect\citeauthoryear{Miller and Mitamura}{Miller and
  Mitamura}{2003}]%
        {miller2003surveys}
\bibfield{author}{\bibinfo{person}{Alan~S Miller} {and} \bibinfo{person}{Tomoko
  Mitamura}.} \bibinfo{year}{2003}\natexlab{}.
\newblock \showarticletitle{Are surveys on trust trustworthy?}
\newblock \bibinfo{journal}{\emph{Social Psychology Quarterly}}
  (\bibinfo{year}{2003}), \bibinfo{pages}{62--70}.
\newblock


\bibitem[\protect\citeauthoryear{Misztal}{Misztal}{2013}]%
        {misztal2013trust}
\bibfield{author}{\bibinfo{person}{Barbara Misztal}.}
  \bibinfo{year}{2013}\natexlab{}.
\newblock \bibinfo{booktitle}{\emph{Trust in modern societies: The search for
  the bases of social order}}.
\newblock \bibinfo{publisher}{John Wiley \& Sons}.
\newblock


\bibitem[\protect\citeauthoryear{Moser, Resnick, and Schoenebeck}{Moser
  et~al\mbox{.}}{2017}]%
        {moser2017community}
\bibfield{author}{\bibinfo{person}{Carol Moser}, \bibinfo{person}{Paul
  Resnick}, {and} \bibinfo{person}{Sarita Schoenebeck}.}
  \bibinfo{year}{2017}\natexlab{}.
\newblock \showarticletitle{Community commerce: Facilitating trust in
  mom-to-mom sale groups on Facebook}. In \bibinfo{booktitle}{\emph{Proceedings
  of the ACM Conference on Human Factors in Computing Systems}}. ACM,
  \bibinfo{pages}{4344--4357}.
\newblock


\bibitem[\protect\citeauthoryear{Nannestad}{Nannestad}{2008}]%
        {nannestad2008have}
\bibfield{author}{\bibinfo{person}{Peter Nannestad}.}
  \bibinfo{year}{2008}\natexlab{}.
\newblock \showarticletitle{What have we learned about generalized trust, if
  anything?}
\newblock \bibinfo{journal}{\emph{Annual Review of Political Science}}
  \bibinfo{volume}{11} (\bibinfo{year}{2008}), \bibinfo{pages}{413--436}.
\newblock


\bibitem[\protect\citeauthoryear{Nyhan}{Nyhan}{2000}]%
        {nyhan2000changing}
\bibfield{author}{\bibinfo{person}{Ronald~C Nyhan}.}
  \bibinfo{year}{2000}\natexlab{}.
\newblock \showarticletitle{Changing the paradigm: Trust and its role in public
  sector organizations}.
\newblock \bibinfo{journal}{\emph{The American Review of Public
  Administration}} \bibinfo{volume}{30}, \bibinfo{number}{1}
  (\bibinfo{year}{2000}), \bibinfo{pages}{87--109}.
\newblock


\bibitem[\protect\citeauthoryear{Paxton}{Paxton}{2007}]%
        {paxton2007association}
\bibfield{author}{\bibinfo{person}{Pamela Paxton}.}
  \bibinfo{year}{2007}\natexlab{}.
\newblock \showarticletitle{Association memberships and generalized trust: A
  multilevel model across 31 countries}.
\newblock \bibinfo{journal}{\emph{Social Forces}} \bibinfo{volume}{86},
  \bibinfo{number}{1} (\bibinfo{year}{2007}), \bibinfo{pages}{47--76}.
\newblock


\bibitem[\protect\citeauthoryear{Perez}{Perez}{2018}]%
        {facebook2018launch}
\bibfield{author}{\bibinfo{person}{Sarah Perez}.}
  \bibinfo{year}{2018}\natexlab{}.
\newblock \bibinfo{title}{Facebook is launching a new Groups tab and plug-in}.
\newblock
  \bibinfo{howpublished}{\url{https://techcrunch.com/2018/05/01/facebook-is-launching-a-new-groups-tab-and-plugin}}.
\newblock
\newblock
\shownote{(Accessed Sep 2018).}


\bibitem[\protect\citeauthoryear{Preece and Maloney-Krichmar}{Preece and
  Maloney-Krichmar}{2005}]%
        {preece2005online}
\bibfield{author}{\bibinfo{person}{Jenny Preece} {and} \bibinfo{person}{Diane
  Maloney-Krichmar}.} \bibinfo{year}{2005}\natexlab{}.
\newblock \showarticletitle{Online communities: Design, theory, and practice}.
\newblock \bibinfo{journal}{\emph{Journal of Computer-Mediated Communication}}
  \bibinfo{volume}{10}, \bibinfo{number}{4} (\bibinfo{year}{2005}).
\newblock


\bibitem[\protect\citeauthoryear{Putnam}{Putnam}{1993}]%
        {putnam1993prosperous}
\bibfield{author}{\bibinfo{person}{Robert~D Putnam}.}
  \bibinfo{year}{1993}\natexlab{}.
\newblock \showarticletitle{The prosperous community}.
\newblock \bibinfo{journal}{\emph{The American Prospect}} \bibinfo{volume}{4},
  \bibinfo{number}{13} (\bibinfo{year}{1993}), \bibinfo{pages}{35--42}.
\newblock


\bibitem[\protect\citeauthoryear{Putnam}{Putnam}{2000}]%
        {putnam2000bowling}
\bibfield{author}{\bibinfo{person}{Robert~D Putnam}.}
  \bibinfo{year}{2000}\natexlab{}.
\newblock \showarticletitle{Bowling alone: America's declining social capital}.
\newblock In \bibinfo{booktitle}{\emph{Culture and Politics}}.
  \bibinfo{publisher}{Springer}, \bibinfo{pages}{223--234}.
\newblock


\bibitem[\protect\citeauthoryear{Qiu, Parigi, and Abrahao}{Qiu
  et~al\mbox{.}}{2018}]%
        {qiu2018more}
\bibfield{author}{\bibinfo{person}{Will Qiu}, \bibinfo{person}{Palo Parigi},
  {and} \bibinfo{person}{Bruno Abrahao}.} \bibinfo{year}{2018}\natexlab{}.
\newblock \showarticletitle{More stars or more reviews?}. In
  \bibinfo{booktitle}{\emph{Proceedings of the ACM Conference on Human Factors
  in Computing Systems}}. ACM, \bibinfo{pages}{153}.
\newblock


\bibitem[\protect\citeauthoryear{Ren, Kraut, and Kiesler}{Ren
  et~al\mbox{.}}{2007}]%
        {ren2007applying}
\bibfield{author}{\bibinfo{person}{Yuqing Ren}, \bibinfo{person}{Robert Kraut},
  {and} \bibinfo{person}{Sara Kiesler}.} \bibinfo{year}{2007}\natexlab{}.
\newblock \showarticletitle{Applying common identity and bond theory to design
  of online communities}.
\newblock \bibinfo{journal}{\emph{Organization Studies}} \bibinfo{volume}{28},
  \bibinfo{number}{3} (\bibinfo{year}{2007}), \bibinfo{pages}{377--408}.
\newblock


\bibitem[\protect\citeauthoryear{Resnick and Zeckhauser}{Resnick and
  Zeckhauser}{2002}]%
        {resnick2002trust}
\bibfield{author}{\bibinfo{person}{Paul Resnick} {and} \bibinfo{person}{Richard
  Zeckhauser}.} \bibinfo{year}{2002}\natexlab{}.
\newblock \showarticletitle{Trust among strangers in Internet transactions:
  Empirical analysis of eBay's reputation system}.
\newblock In \bibinfo{booktitle}{\emph{The Economics of the Internet and
  E-commerce}}. \bibinfo{publisher}{Emerald Group Publishing Limited},
  \bibinfo{pages}{127--157}.
\newblock


\bibitem[\protect\citeauthoryear{Ridings, Gefen, and Arinze}{Ridings
  et~al\mbox{.}}{2002}]%
        {ridings2002some}
\bibfield{author}{\bibinfo{person}{Catherine~M Ridings}, \bibinfo{person}{David
  Gefen}, {and} \bibinfo{person}{Bay Arinze}.} \bibinfo{year}{2002}\natexlab{}.
\newblock \showarticletitle{Some antecedents and effects of trust in virtual
  communities}.
\newblock \bibinfo{journal}{\emph{The Journal of Strategic Information
  Systems}} \bibinfo{volume}{11}, \bibinfo{number}{3-4} (\bibinfo{year}{2002}),
  \bibinfo{pages}{271--295}.
\newblock


\bibitem[\protect\citeauthoryear{Rotter}{Rotter}{1971}]%
        {rotter1971generalized}
\bibfield{author}{\bibinfo{person}{Julian~B Rotter}.}
  \bibinfo{year}{1971}\natexlab{}.
\newblock \showarticletitle{Generalized expectancies for interpersonal trust.}
\newblock \bibinfo{journal}{\emph{American Psychologist}} \bibinfo{volume}{26},
  \bibinfo{number}{5} (\bibinfo{year}{1971}), \bibinfo{pages}{443}.
\newblock


\bibitem[\protect\citeauthoryear{Rousseau, Sitkin, Burt, and Camerer}{Rousseau
  et~al\mbox{.}}{1998}]%
        {rousseau1998not}
\bibfield{author}{\bibinfo{person}{Denise~M Rousseau}, \bibinfo{person}{Sim~B
  Sitkin}, \bibinfo{person}{Ronald~S Burt}, {and} \bibinfo{person}{Colin
  Camerer}.} \bibinfo{year}{1998}\natexlab{}.
\newblock \showarticletitle{Not so different after all: A cross-discipline view
  of trust}.
\newblock \bibinfo{journal}{\emph{Academy of Management Review}}
  \bibinfo{volume}{23}, \bibinfo{number}{3} (\bibinfo{year}{1998}),
  \bibinfo{pages}{393--404}.
\newblock


\bibitem[\protect\citeauthoryear{Schoorman, Mayer, and Davis}{Schoorman
  et~al\mbox{.}}{2007}]%
        {schoorman2007integrative}
\bibfield{author}{\bibinfo{person}{F~David Schoorman}, \bibinfo{person}{Roger~C
  Mayer}, {and} \bibinfo{person}{James~H Davis}.}
  \bibinfo{year}{2007}\natexlab{}.
\newblock \showarticletitle{An integrative model of organizational trust: Past,
  present, and future}.
\newblock \bibinfo{journal}{\emph{Academy of Management Review}}
  \bibinfo{volume}{32}, \bibinfo{number}{2} (\bibinfo{year}{2007}).
\newblock


\bibitem[\protect\citeauthoryear{Simon and Brown}{Simon and Brown}{1987}]%
        {simon1987perceived}
\bibfield{author}{\bibinfo{person}{Bernd Simon} {and} \bibinfo{person}{Rupert
  Brown}.} \bibinfo{year}{1987}\natexlab{}.
\newblock \showarticletitle{Perceived intragroup homogeneity in
  minority-majority contexts.}
\newblock \bibinfo{journal}{\emph{Journal of Personality and Social
  Psychology}} (\bibinfo{year}{1987}).
\newblock


\bibitem[\protect\citeauthoryear{Six and Sorge}{Six and Sorge}{2008}]%
        {six2008creating}
\bibfield{author}{\bibinfo{person}{Fr{\'e}d{\'e}rique Six} {and}
  \bibinfo{person}{Arndt Sorge}.} \bibinfo{year}{2008}\natexlab{}.
\newblock \showarticletitle{Creating a high-trust organization: An exploration
  into organizational policies that stimulate interpersonal trust building}.
\newblock \bibinfo{journal}{\emph{Journal of Management Studies}}
  \bibinfo{volume}{45}, \bibinfo{number}{5} (\bibinfo{year}{2008}),
  \bibinfo{pages}{857--884}.
\newblock


\bibitem[\protect\citeauthoryear{Stokes}{Stokes}{1983}]%
        {stokes1983components}
\bibfield{author}{\bibinfo{person}{Joseph~Powell Stokes}.}
  \bibinfo{year}{1983}\natexlab{}.
\newblock \showarticletitle{Components of group cohesion: Intermember
  attraction, instrumental value, and risk taking}.
\newblock \bibinfo{journal}{\emph{Small Group Behavior}}
  (\bibinfo{year}{1983}).
\newblock


\bibitem[\protect\citeauthoryear{Survey}{Survey}{2018}]%
        {wvs}
\bibfield{author}{\bibinfo{person}{World~Values Survey}.}
  \bibinfo{year}{Accessed Aug 2018}\natexlab{}.
\newblock \bibinfo{title}{World Values Survey database}.
\newblock
\newblock
\urldef\tempurl%
\url{http://www.worldvaluessurvey.org/WVSContents.jsp}
\showURL{%
\tempurl}


\bibitem[\protect\citeauthoryear{Taylor, Funk, and Clark}{Taylor
  et~al\mbox{.}}{2007}]%
        {taylor2007americans}
\bibfield{author}{\bibinfo{person}{Paul Taylor}, \bibinfo{person}{Cary Funk},
  {and} \bibinfo{person}{April Clark}.} \bibinfo{year}{2007}\natexlab{}.
\newblock \showarticletitle{Americans and social trust: Who, where and why}.
\newblock \bibinfo{journal}{\emph{A Social Trends Report}}
  (\bibinfo{year}{2007}).
\newblock


\bibitem[\protect\citeauthoryear{Ugander, Backstrom, Marlow, and
  Kleinberg}{Ugander et~al\mbox{.}}{2012}]%
        {ugander2012structural}
\bibfield{author}{\bibinfo{person}{Johan Ugander}, \bibinfo{person}{Lars
  Backstrom}, \bibinfo{person}{Cameron Marlow}, {and} \bibinfo{person}{Jon
  Kleinberg}.} \bibinfo{year}{2012}\natexlab{}.
\newblock \showarticletitle{Structural diversity in social contagion}.
\newblock \bibinfo{journal}{\emph{Proceedings of the National Academy of
  Sciences}} (\bibinfo{year}{2012}), \bibinfo{pages}{5962--5966}.
\newblock


\bibitem[\protect\citeauthoryear{Uzzi}{Uzzi}{1996}]%
        {uzzi1996sources}
\bibfield{author}{\bibinfo{person}{Brian Uzzi}.}
  \bibinfo{year}{1996}\natexlab{}.
\newblock \showarticletitle{The sources and consequences of embeddedness for
  the economic performance of organizations: The network effect}.
\newblock \bibinfo{journal}{\emph{American Sociological Review}}
  (\bibinfo{year}{1996}), \bibinfo{pages}{674--698}.
\newblock


\bibitem[\protect\citeauthoryear{Van~Vugt and Hart}{Van~Vugt and Hart}{2004}]%
        {van2004social}
\bibfield{author}{\bibinfo{person}{Mark Van~Vugt} {and}
  \bibinfo{person}{Claire~M Hart}.} \bibinfo{year}{2004}\natexlab{}.
\newblock \showarticletitle{Social identity as social glue: the origins of
  group loyalty.}
\newblock \bibinfo{journal}{\emph{Journal of Personality and Social
  Psychology}} \bibinfo{volume}{86}, \bibinfo{number}{4}
  (\bibinfo{year}{2004}), \bibinfo{pages}{585}.
\newblock


\bibitem[\protect\citeauthoryear{Vigoda-Gadot and Talmud}{Vigoda-Gadot and
  Talmud}{2010}]%
        {vigoda2010organizational}
\bibfield{author}{\bibinfo{person}{Eran Vigoda-Gadot} {and}
  \bibinfo{person}{Ilan Talmud}.} \bibinfo{year}{2010}\natexlab{}.
\newblock \showarticletitle{Organizational politics and job outcomes: The
  moderating effect of trust and social support}.
\newblock \bibinfo{journal}{\emph{Journal of Applied Social Psychology}}
  \bibinfo{volume}{40}, \bibinfo{number}{11} (\bibinfo{year}{2010}),
  \bibinfo{pages}{2829--2861}.
\newblock


\bibitem[\protect\citeauthoryear{Walther and Bunz}{Walther and Bunz}{2005}]%
        {walther2005rules}
\bibfield{author}{\bibinfo{person}{Joseph~B Walther} {and}
  \bibinfo{person}{Ulla Bunz}.} \bibinfo{year}{2005}\natexlab{}.
\newblock \showarticletitle{The rules of virtual groups: Trust, liking, and
  performance in computer-mediated communication}.
\newblock \bibinfo{journal}{\emph{Journal of Communication}}
  \bibinfo{volume}{55}, \bibinfo{number}{4} (\bibinfo{year}{2005}),
  \bibinfo{pages}{828--846}.
\newblock


\bibitem[\protect\citeauthoryear{Williamson}{Williamson}{1993}]%
        {williamson1993calculativeness}
\bibfield{author}{\bibinfo{person}{Oliver~E Williamson}.}
  \bibinfo{year}{1993}\natexlab{}.
\newblock \showarticletitle{Calculativeness, trust, and economic organization}.
\newblock \bibinfo{journal}{\emph{Journal of Law and Economics}}
  \bibinfo{volume}{36}, \bibinfo{number}{1, Part 2} (\bibinfo{year}{1993}),
  \bibinfo{pages}{453--486}.
\newblock


\bibitem[\protect\citeauthoryear{Wu, Hu, and Wu}{Wu et~al\mbox{.}}{2010}]%
        {wu2010effects}
\bibfield{author}{\bibinfo{person}{Guohua Wu}, \bibinfo{person}{Xiaorui Hu},
  {and} \bibinfo{person}{Yuhong Wu}.} \bibinfo{year}{2010}\natexlab{}.
\newblock \showarticletitle{Effects of perceived interactivity, perceived web
  assurance and disposition to trust on initial online trust}.
\newblock \bibinfo{journal}{\emph{Journal of Computer-Mediated Communication}}
  \bibinfo{volume}{16}, \bibinfo{number}{1} (\bibinfo{year}{2010}),
  \bibinfo{pages}{1--26}.
\newblock


\bibitem[\protect\citeauthoryear{Yakovleva, Reilly, and Werko}{Yakovleva
  et~al\mbox{.}}{2010}]%
        {yakovleva2010we}
\bibfield{author}{\bibinfo{person}{Maria Yakovleva}, \bibinfo{person}{Richard~R
  Reilly}, {and} \bibinfo{person}{Robert Werko}.}
  \bibinfo{year}{2010}\natexlab{}.
\newblock \showarticletitle{Why do we trust? Moving beyond individual to dyadic
  perceptions.}
\newblock \bibinfo{journal}{\emph{Journal of Applied Psychology}}
  \bibinfo{volume}{95}, \bibinfo{number}{1} (\bibinfo{year}{2010}),
  \bibinfo{pages}{79}.
\newblock


\bibitem[\protect\citeauthoryear{Yamagishi, Matsuda, Yoshikai, Takahashi, and
  Usui}{Yamagishi et~al\mbox{.}}{2009}]%
        {yamagishi2004solving}
\bibfield{author}{\bibinfo{person}{Toshio Yamagishi}, \bibinfo{person}{Masafumi
  Matsuda}, \bibinfo{person}{Noriaki Yoshikai}, \bibinfo{person}{Hiroyuki
  Takahashi}, {and} \bibinfo{person}{Yukihiro Usui}.}
  \bibinfo{year}{2009}\natexlab{}.
\newblock \showarticletitle{Solving the lemons problem with reputation}.
\newblock In \bibinfo{booktitle}{\emph{eTrust: Forming Relationships in the
  Online World}}.
\newblock


\bibitem[\protect\citeauthoryear{Yuki, Maddux, Brewer, and Takemura}{Yuki
  et~al\mbox{.}}{2005}]%
        {yuki2005cross}
\bibfield{author}{\bibinfo{person}{Masaki Yuki}, \bibinfo{person}{William~W
  Maddux}, \bibinfo{person}{Marilynn~B Brewer}, {and} \bibinfo{person}{Kosuke
  Takemura}.} \bibinfo{year}{2005}\natexlab{}.
\newblock \showarticletitle{Cross-cultural differences in relationship-and
  group-based trust}.
\newblock \bibinfo{journal}{\emph{Personality and Social Psychology Bulletin}}
  \bibinfo{volume}{31}, \bibinfo{number}{1} (\bibinfo{year}{2005}),
  \bibinfo{pages}{48--62}.
\newblock


\bibitem[\protect\citeauthoryear{Zelmer}{Zelmer}{2003}]%
        {zelmer2003linear}
\bibfield{author}{\bibinfo{person}{Jennifer Zelmer}.}
  \bibinfo{year}{2003}\natexlab{}.
\newblock \showarticletitle{Linear public goods experiments: A meta-analysis}.
\newblock \bibinfo{journal}{\emph{Experimental Economics}} \bibinfo{volume}{6},
  \bibinfo{number}{3} (\bibinfo{year}{2003}), \bibinfo{pages}{299--310}.
\newblock


\bibitem[\protect\citeauthoryear{Zhao, Lu, Wang, Chau, and Zhang}{Zhao
  et~al\mbox{.}}{2012}]%
        {zhao2012cultivating}
\bibfield{author}{\bibinfo{person}{Ling Zhao}, \bibinfo{person}{Yaobin Lu},
  \bibinfo{person}{Bin Wang}, \bibinfo{person}{Patrick~YK Chau}, {and}
  \bibinfo{person}{Long Zhang}.} \bibinfo{year}{2012}\natexlab{}.
\newblock \showarticletitle{Cultivating the sense of belonging and motivating
  user participation in virtual communities: A social capital perspective}.
\newblock \bibinfo{journal}{\emph{International Journal of Information
  Management}} \bibinfo{volume}{32}, \bibinfo{number}{6}
  (\bibinfo{year}{2012}), \bibinfo{pages}{574--588}.
\newblock


\end{thebibliography}

\end{document}